\documentclass[aps,prl,twocolumn,showpacs,amsmath,amssymb,superscriptaddress]{revtex4}
\usepackage{graphicx}
\usepackage{dcolumn}
\usepackage{bm}
\usepackage{natbib}
\usepackage{amsmath}
\usepackage{array}

\begin{document}
\title{Hydrodynamic Correlation Functions of a Driven Granular Fluid in Steady State}

\author{Katharina Vollmayr-Lee}
 \email{kvollmay@bucknell.edu}
\affiliation{Department of Physics and Astronomy, Bucknell University,
      Lewisburg, Pennsylvania 17837, USA}
\author{Timo Aspelmeier}
\affiliation{Max-Planck-Institut f\"ur Dynamik und Selbstorganisation, Bunsenstr. 10,
37073 G\"ottingen, Germany}
\affiliation{Scivis GmbH, Bertha-von-Suttner-Str. 5, 37085 G\"ottingen, Germany}
\author{Annette Zippelius}
\affiliation{Max-Planck-Institut f\"ur Dynamik und Selbstorganisation, Bunsenstr. 10,
37073 G\"ottingen, Germany}
\affiliation{Georg-August-Universit\"at G\"ottingen, 
Institut f\"ur Theoretische Physik, 
Friedrich-Hund-Platz 1, 37077 G\"ottingen, Germany}

%\date{July 21, 2010}
\date{\today}
\begin{abstract}
  We study a homogeneously driven granular fluid of hard spheres at
  intermediate volume fractions and focus on time-delayed correlation
  functions in the stationary state. Inelastic collisions are modeled
  by incomplete normal restitution, allowing for efficient simulations
  with an event-driven algorithm. The incoherent scattering function,
  $F_{\rm incoh}(q,t)$, is seen to follow time-density superposition
  with a relaxation time that increases significantly as volume
  fraction increases. The statistics of particle displacements is
  approximately Gaussian. For the coherent scattering function
  $S(q,\omega)$ we compare our results to the predictions of
  generalized fluctuating hydrodynamics which takes into account that
  temperature fluctuations decay either diffusively or with a finite relaxation rate,
  depending on wave number and inelasticity. For sufficiently small
  wave number $q$ we observe sound waves in the coherent scattering
  function $S(q,\omega)$ and the longitudinal current correlation 
  function $C_{\mathrm l}(q,\omega)$. We determine the speed of sound and the
  transport coefficients and compare them to the results of kinetic
  theory.
\end{abstract}

% suggestion
%61.20.Lc 	Time-dependent properties; relaxation (for glass transitions, see 64.70.P-)
%51.20.+d       Viscosity, diffusion, and thermal conductivity
%45.70.-n 	Granular systems 
%47.57.Gc 	Granular flow 

\maketitle
\section{I Introduction}

The long wavelength, low frequency dynamics of granular fluids is
frequently described by phenomenological hydrodynamic equations
\cite{campbell90,brey98,sela98,goldhirsch03}. In contrast to a fluid
composed of elastically colliding particles, the total energy of the
system is not conserved, implying a finite decay rate of the
temperature in the limit of long wavelength. Hence, strictly speaking, 
the temperature is not a hydrodynamic variable.  More generally, the
scale separation required by hydrodynamics has been questioned
\cite{tan98}. If the system is not driven, the homogeneous state is
unstable \cite{goldhirsch93} and large spatial gradients develop ---
invalidating a hydrodynamic approach. A third point of criticism 
refers to the pressure in the Navier-Stokes equation. Closure of the
hydrodynamic equations requires an equation of state to express the
pressure \cite{herbst04} in terms of density and temperature. However,
an equation of state is expected to exist only in an equilibrium
state.  Given these problems, the hydrodynamic approach has been
mainly restricted to small inelasticity, such that the decay rate of
the temperature is small, the time for the build-up of spatial
inhomogeneities is long, and an equation of state is approximately
valid. In this limit, kinetic theory has provided a basis for the
hydrodynamic equations and given explicit expressions for the
transport coefficients \cite{jenkins83,lun84,poeschel}.

Driving a granular fluid allows for a compensation of the energy which
is dissipated in collisions, such that a non-equilibrium stationary
state (NESS) is reached. In experiment, the driving is frequently
performed by shearing \cite{lutsko2001,santos2008}, with vibrating
walls \cite{olafsen1999,reyes2008,brey2010} or by driving the system
homogeneously \cite{schroeter05,durian06}. To test a hydrodynamic
approach in the NESS we want to avoid new length scales, which might
be generated by driving through the boundaries, when the agitation
decays over a characteristic length, e.g. the width of a shear band.
Hence in the following, we consider a homogeneously driven granular
fluid \cite{noije1998,noije1999,pagonabarraga2001,fiege2009,kranz2010}
and set out to investigate the validity of the hydrodynamic approach in
the NESS. 

We present in this paper results for  a homogeneously driven 
%(while conserving total momentum globally) 
system of hard spheres with moderate inelasticity, parametrized by 
a coefficient of restitution $\epsilon=0.8,
0.9$ and $1.0$ (elastic). % at low to intermediate volume fractions $0.05
%\le \eta \le 0.4$.
We use event-driven simulations to focus on the dynamics of the
system.  Previous studies of correlation functions for granular fluids
have been either on density-density correlations at the same time,
such as the structure factor $S(q)$ \cite{noije1998,noije1999} and the
pair correlation function \cite{panaitescu2010}, %otsuki2009?
or on velocity-velocity correlations \cite{fiege2009} (and references
therein).  We focus here on time dependent spatial correlations
\cite{kranz2010} at  volume fraction $0.05\leq
\eta \leq 0.4$ and compute the incoherent and coherent intermediate
scattering functions.

% with the intend to compare with the predictions of Noije et
%al. \cite{noije1999}  using fluctuating hydrodynamics.
The former entails information about the motion of a tagged particle,
which is expected to be diffusive at long times.  We find that the
incoherent scattering function is well approximated by a Gaussian and
obeys time-density superposition. The divergence of the relaxation
time as a function of $\eta$ occurs not only for the elastic case but
also for the inelastic case, consistent with the results of Reyes et
al. \cite{reyes2008} and \cite{kranz2010}.

The coherent correlations reveal the collective dynamics of the fluid:
damped sound waves and relaxation of temperature fluctuations.  We
determine the dynamic structure factor $S(q,\omega)$ and compare our
data quantitatively to the predictions of van Noije et al. \cite{noije1999}
using fluctuating hydrodynamics. The agreement between simulations
and theory is quite good: damped sound waves are indeed
observed for small wave numbers $q$ and the velocity and
damping of sound can be determined. Temperature
fluctuations are found to decay either diffusively or with a finite
rate, depending on $q$. The transport coefficients are compared to the
predictions of kinetic theory and found to agree well.

In the following we specify model and simulation details in Sec.\ II.
%\ref{sec:model}. 
Subsequently, in Sec.\ III we discuss the incoherent scattering
function, the mean square displacement, and the diffusion
constant. Data for the intermediate coherent scattering function 
and longitudinal current correlation function 
are presented in
Sec.\ IV. A hydrodynamic model, which was first
introduced in Ref.~\cite{noije1998}, is discussed in Sec.\ V and
compared to the
simulation data for the coherent scattering function in Sec.\ VI.

%the hydrodynamic model in Sec.\ V, %\ref{sec:hydrodynamics}, 
%and results in Sec.\ III, %\ref{sec:Fs}, 
%Sec.\ IV %\ref{sec:Fqt} 
%and Sec.\ VI. %\ref{sec:Fqomega}.

\section{II Model and Simulation Details}
\label{sec:model}
We investigate a system of $N$ monodisperse
hard spheres of diameter $a$ and mass $m$ at volume fraction 
$\eta=\frac{N\pi a^3}{6V}$. The time
evolution is governed by instantaneous inelastic two-particle collisions.
We consider here only the simplest model of an inelastic two-body
collision, described by incomplete normal restitution. The change of
the relative velocity $\mathbf g := \mathbf v_1 - \mathbf v_2$ of the
two colliding particles is given by
\begin{equation}
  \left( \mathbf g \cdot \mathbf n \right)^\prime =
 -\varepsilon \left( \mathbf g \cdot \mathbf n \right),
\end{equation}
where primed quantities indicate post-collisional velocities and
unprimed ones refer to precollisional ones. The unit vector $\mathbf n
:= (\mathbf r_1 - \mathbf r_2)/\left\vert (\mathbf r_1 - \mathbf r_2)
\right\vert$ connects the centers of the two spheres, and
$\varepsilon = $const. $\in\left[0,1\right]$ denotes the coefficient of normal
restitution, with $\varepsilon=1.0$ in the elastic limit. 
%Annette: Brilliantov Eq.(5.2) (1-eps^2)mg^2/4 and for average energy  ?
%    Do we need to specify Delta E here?
%The energy loss is given by $\Delta E=(1-\varepsilon^2)
%mg^2/8$ with $\varepsilon=1$ in the elastic limit. 
The postcollisional
velocities of the two colliding spheres are given by 
\begin{align}
{\mathbf
  v}_1^\prime = \mathbf v_1 - \frac{\left ( 1+\varepsilon \right )}{2} (
\mathbf n \cdot \mathbf g ) \mathbf n \\
{\mathbf
  v}_2^\prime=\mathbf v_2 +  \frac{\left ( 1+\varepsilon \right )}{2} (
\mathbf n \cdot \mathbf g ) \mathbf n.
\end{align}

Due to the inelastic nature of the collisions, we have to feed energy
into the system in order to maintain a stationary state. 
The simplest bulk
driving~\cite{MacKintosh} consists of a kick of a given particle, say particle
$i$, instantaneously at time $t$, which corresponds to
\begin{equation}
 \mathbf v_i (t) = \mathbf v_i(t_o) +\int_{t_0}^tds\bm\xi_i(s).
\end{equation}
The noise $\bm\xi_i(t)$ is Gaussian with zero mean and variance
\begin{equation}
\langle\xi_i^{(\alpha)}(t)
\xi_j^{(\beta)}(t^\prime)\rangle=\xi_0^2 \delta_{i,j} \delta_{\alpha\beta}
\delta (t-t^\prime)
\end{equation}
for the cartesian components $\xi_i^{(\alpha)}$, $\alpha=x,y,z$. The
stochastic process is implemented in the simulation by 
kicking the particles randomly with amplitude $v_\mathrm{Dr}$ and 
frequency $f_\mathrm{Dr}$.

If a single particle is kicked at a particular instant, momentum is
not conserved. Due to the random direction of the kicks the time
average will restore the conservation of global momentum, but only on
average. Momentum conservation is known to be essential for the
dynamic correlation functions in the limit of long wavelength and long
times.  Hence we choose a driving mechanism in which pairs of
particles are kicked in opposite directions~\cite{dissip_part}. The
pairs are fixed globally so that the total momentum is conserved at
each instant of time. Denoting the partner of particle $i$ by $p(i)$,
the random force correlation is given by
%for xi see (15) Noije
\begin{equation}
\langle\xi_i^{(\alpha)}(t)
\xi_j^{(\beta)}(t^\prime)\rangle=\xi_0^2(\delta_{j,i}-\delta_{j,p(i)}) \delta_{\alpha\beta}
\delta (t-t^\prime).
\end{equation}
It is also possible to ensure momentum
conservation on small scales by choosing pairs of neighboring
particles and by kicking them in opposite directions. However this is not
pursued here.

For the event-driven simulations we use the optimized algorithm of 
Lubachevsky \cite{lubachevsky1991} adapted to granular media
\cite{fiege2009}. To avoid the inelastic collapse 
we use the technique of virtual hulls around the spheres as described 
in \cite{fiege2009}. Particles are colliding elastically when they 
are a diameter $a$ apart and the dissipation takes place when the 
colliding spheres are receding and separated by $\left (1+10^{-4}\right ) a$.
    %To simulate at constant temperature $T$, $v_\mathrm{Dr}$ 
    %and $f_\mathrm{Dr}$ were chosen such that 
    %$v_\mathrm{Dr}^2 f_\mathrm{Dr}=\xi_0^2$
    %=\frac{(1-\varepsilon^2)}{3m} \omega_\mathrm{E} T$,
    %where $\omega_\mathrm{E}$ is the angular Enskog frequency
    %\begin{equation}
    %\label{eq:omegaE}
    %\omega_\mathrm{E} = 4 \frac{N}{V} d^2 g(a) \sqrt{\frac{\pi T}{m}}
    %\end{equation}
    %with the Carnahan-Startling pair correlation function at contact 
    %\begin{equation}
    %\label{eq:g2}
    %g(d) = \frac{\left (1 - \eta/2 \right )}{\left (1-\eta \right )^3}
    %\end{equation}
%Noije (38)
With the appropriate choice of $v_\mathrm{Dr}^2 f_\mathrm{Dr}=\xi_0^2$ we ensured 
constant temperature and in all following
we chose units such that $m=a=T=1$. 
All simulations were with a cubic box and periodic boundary conditions.
The simulation results of Sec.\ III %\ref{sec:Fs} 
are for $N=200000$ and two independent simulation 
runs \footnote{Only exception are the results of Fig.~\ref{fig:Gaussian}
for which we used five independent simulation runs.},
whereas for the results of Sec.\ IV %\ref{sec:Fqt} 
and Sec.\ VI %\ref{sec:Fqomega} 
we needed more statistics and 
therefore used $N=10000$ and 100 independent simulation runs.
In each set of simulations we first equilibrated at $\varepsilon=1.0$  at the 
desired volume fraction, followed by a relaxation to a stationary
state at $\varepsilon \ne 1$ (achieved with a simulation run of at least 
100 time units) and consecutive production runs.
Independent configurations were taken from the initial elastic equilibration run
separated in time by at least 1000 time units.
%in the case of $N=200000$  for $2500/1000/400(/850/400)$ time units and 
%in the case of  $N=10000$ for $4000/2000/1000/$ time units for
%$\eta=0.05/0.1/0.2(/0.3/0.4)$  respectively. Relaxation to a
%stationary state  with $\varepsilon \ne 1$
%was achieved with $N=200000$ in 4000/1800/1200(/600/300) time units and with $N=10000$ in
%120/117/98 time units for $\eta=0.05/0.1/0.2(/0.3/0.4)$ respectively.

\section{III Intermediate Incoherent Scattering Function and Self Diffusion Constant}
\label{sec:Fs}

\begin{figure}
\includegraphics[width=0.47\textwidth]{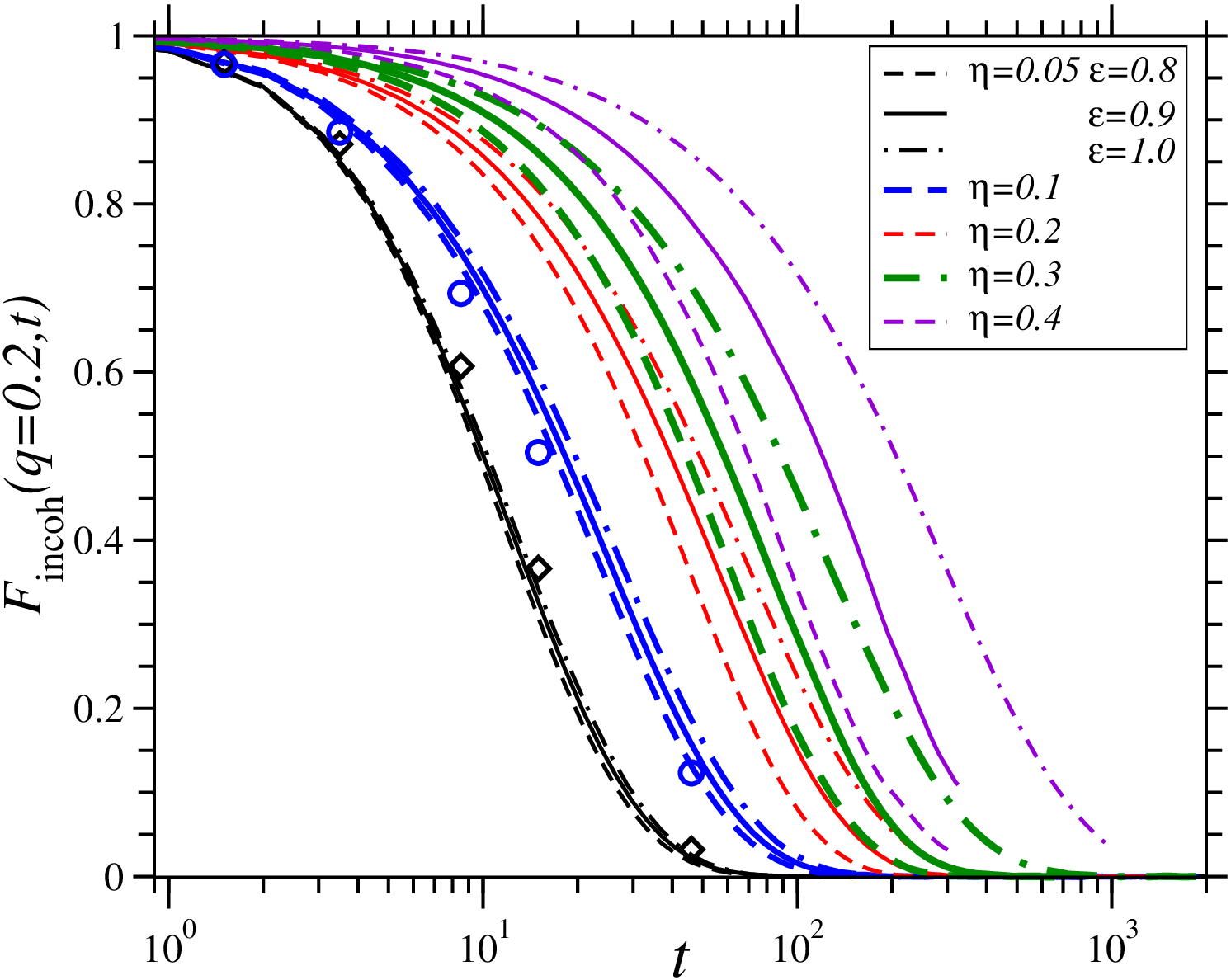}
\caption{Incoherent intermediate scattering function for several values of volume
  fraction $\eta$ and coefficient of restitution $\varepsilon$. All lines for 
   $N=200000$ where $\varepsilon=0.8/0.9/1.0$ are indicated with 
dashed/solid/dot-dashed lines respectively. $\eta=0.05$ corresponds to the left 
and $\eta=0.4$ to the right lines. All error bars are of the order of $10^{-3}$.
Open diamonds and circles are for $N=10000$, $\varepsilon=0.8$ and $\eta=0.05,0.1$ 
respectively.}
\label{fig:fsqt}
\end{figure}

\begin{figure}
\includegraphics[width=0.47\textwidth]{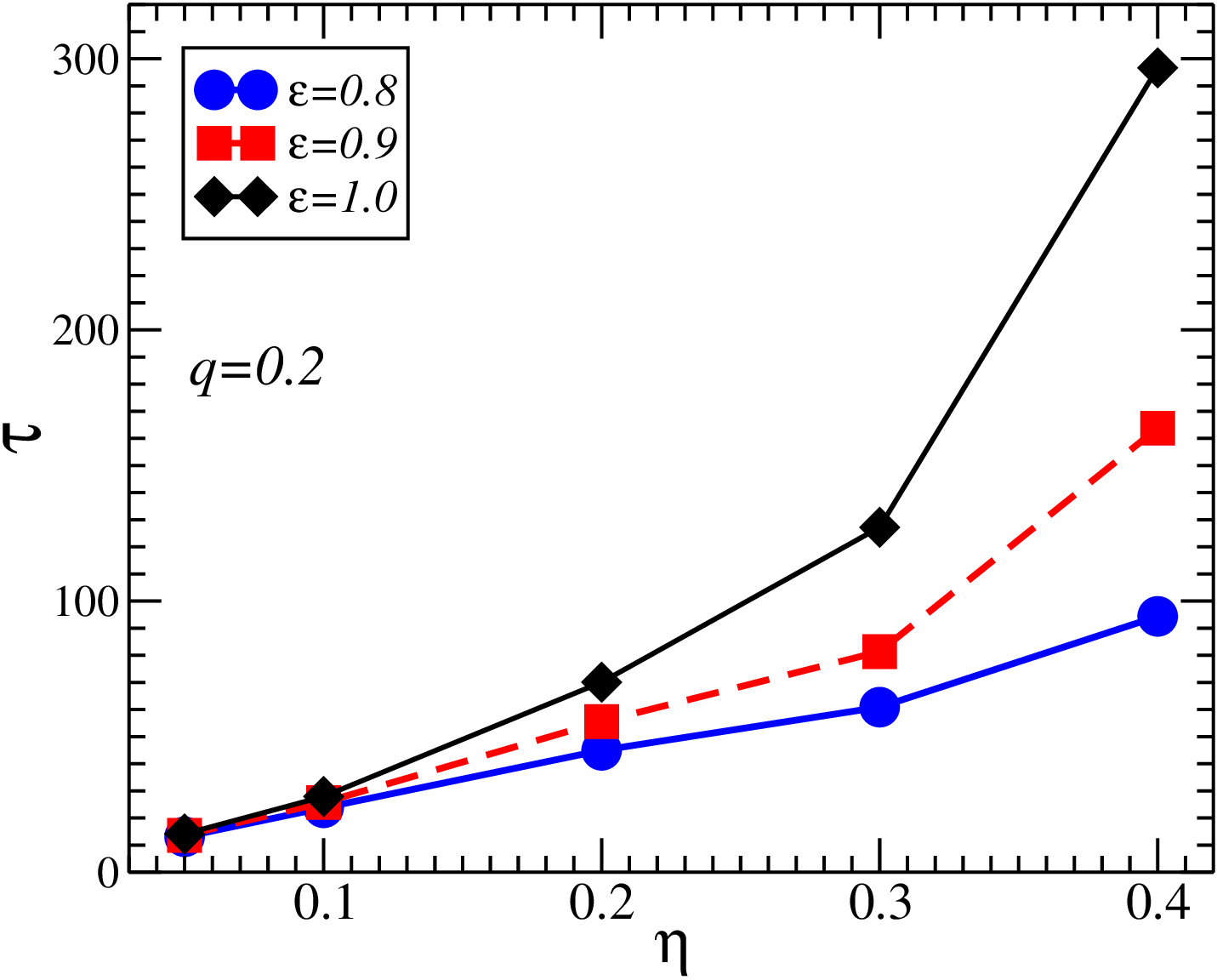}
\caption{Relaxation time of the incoherent scattering function as a
  function of volume
  fraction $\eta$ for several values of $\varepsilon$ for 
  simulation runs with $N=200000$ particles.
%  Straight lines correspond to power law fits 
%$\tau=\tau_0 (\eta_{\rm c}-\eta)^{-\gamma}$ where 
%$\tau_0=\tau(\eta=0.4)/\left ((\eta_{\rm c}-0.4)^{-\gamma}\right )$ and 
%%$\gamma=2.58$ and $\eta_{\rm c}=0.64056/0.716281/0.869646$ for $\varepsilon=1.0/0.9/0.8$.
%$\gamma=2.58$ and $\eta_{\rm c}=0.641/0.716/0.870$ for $\varepsilon=1.0/0.9/0.8$.
%We chose $\gamma=2.58$ as determined by Barrat et al. \cite{barrat1989} for 
%elastic hard spheres. 
%Dashed lines correspond to Vogel-Fulcher fits 
%$\tau=\tau_0 \exp \left (B/(\eta_{\rm c}-\eta) \right )$ where $\tau_0$ was
%chosen such that $\tau(0.4)$ is equal to the simulation value, $\eta_{\rm c}$ 
%%was the same as for the power law fit and $B=0.836912/1.05514/1.48967$ for 
%was the same as for the power law fit and $B=0.837/1.055/1.490$ for 
%$\varepsilon=1.0/0.9/0.8$.
%Variation of $\eta_{\rm c}, \gamma, B$ did not improve the quality of the fits
%and in all cases the power law fit was superior to the Vogel-Fulcher fit.
%(Annette: only if only eta=0.2,0.3 are fit, then the fit is better for this small 
%eta-range, but improvement for one parameter-fit to 2 points is not telling us much.)
 }
\label{fig:tau}
\end{figure}

In this section we investigate time delayed correlations of a 
single tagged particle.
In Fig.~\ref{fig:fsqt} we show for volume fractions
$0.05\leq\eta\leq 0.4$ and for inelasticities
 $\varepsilon =0.8,0.9,1.0$ the incoherent intermediate scattering function
\begin{equation}
F_\mathrm{incoh}(\mathbf q,t)=\left \langle \frac{1}{N}\sum_{i=1}^N
e^{i\mathbf q \cdot (\mathbf r_i(t)-\mathbf r_i(0)}\right \rangle.
\label{eq:fqtdef}
\end{equation}
Since $F_\mathrm{incoh}(\mathbf q,t)$ is a measure of the correlation
of particle $i$ at position $\mathbf r_i(t)$ at time $t$ and at 
position $\mathbf r_i(0)$ at time $t=0$, we find 
as expected that $F_\mathrm{incoh}(\mathbf q,t)$ decreases with increasing time.
%with diffusive behavior for long times due to particle number conservation. 
With decreasing densities the $\mathbf r_i(t)$ and $\mathbf r_i(0)$
become more quickly uncorrelated and therefore the decay is faster for 
smaller volume fractions.
For the lowest densities the inelastic
 system can hardly be distinguished from the elastic case. For higher
 densities the relaxation is increasingly faster for the more
 inelastic systems. To quantify this effect we plot in Fig.~\ref{fig:tau} 
the relaxation time
$\tau$ when the incoherent intermediate scattering function 
has decayed to $1/e$ of its initial
 value, i.e. $F_\mathrm{incoh}(\mathbf q,\tau)=1/e$. 
Clearly the elastic system shows the most rapid increase of
 relaxation time with density, even though the highest volume fraction
 ($\eta=0.4$) is still well below the critical value for the glass
 transition.
% ($\eta_{\mathrm{glass}}=0.6$ 
%\cite{barrat1989,voigtmann2004,vanMegen1991,mason1995}.) 
% also W.K. Kegel and A. van Blaaderen, Science 287, 290 (2000)
The slowing down is weaker for the
 inelastic systems. However, the inelastic system also 
 shows an increase by a factor of 12 ($\varepsilon=0.9$) and 7
 ($\varepsilon=0.8$). This indication for a precursor of a glass transition even
 for the inelastic system is consistent with the higher density results 
 of Kranz et al. \cite{kranz2010} (theory and simulation) and of
Reis et al. \cite{reis2007}  and Reyes et al. \cite{reyes2008} (experiment.)
 
\begin{figure}
\includegraphics[width=0.47\textwidth]{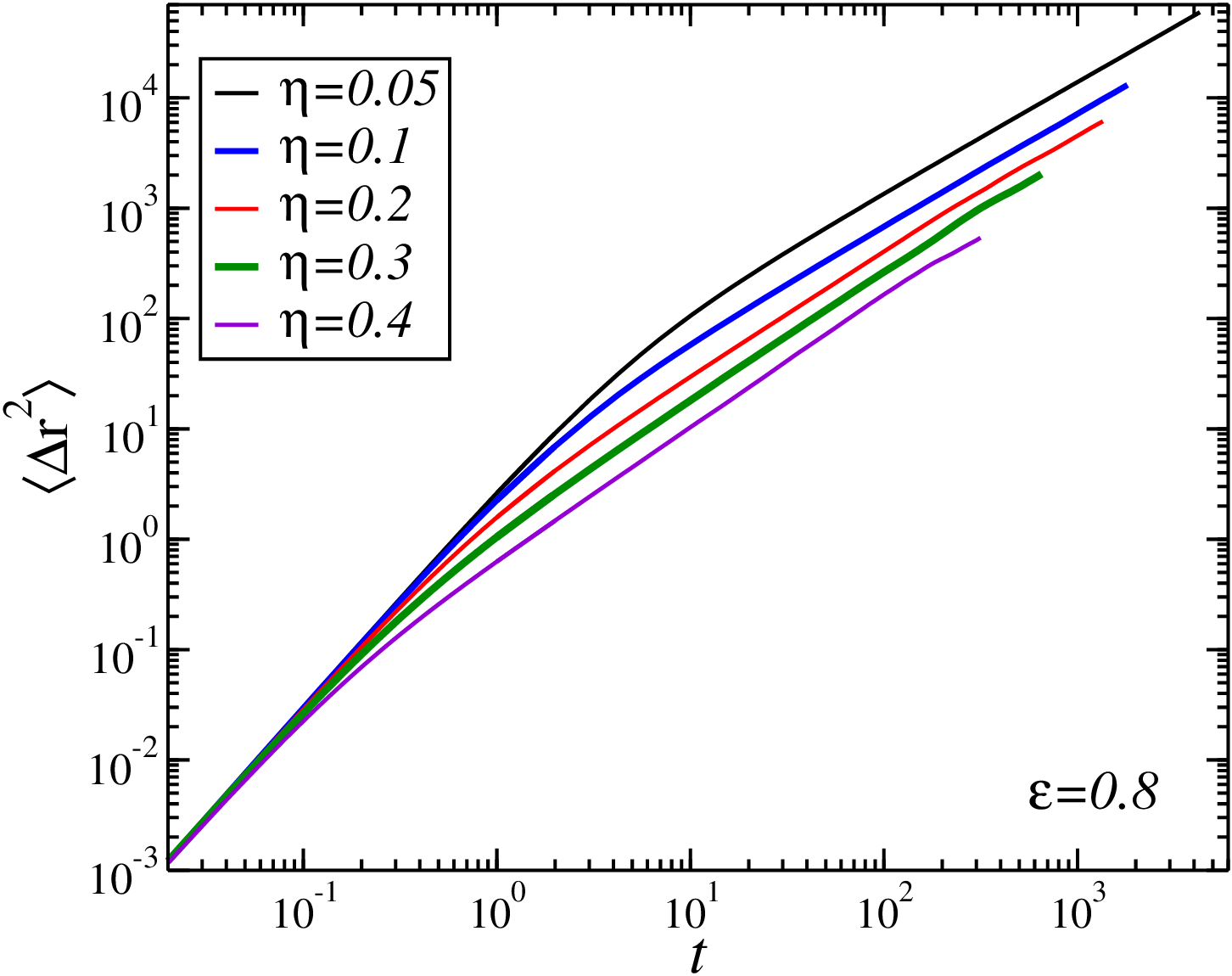}
\caption{Mean square displacement for $\varepsilon=0.8$ and various volume fractions.}
\label{fig:msdeps08}
\end{figure}

 The intermediate incoherent scattering function, $F_\mathrm{incoh}(\mathbf q,t)$,
 is often approximated by a Gaussian
\begin{equation}
F_\mathrm{incoh}(\mathbf q,t)=e^{\frac{-q^2}{6}\left \langle \Delta r^2(t) \right \rangle},
\label{Eq:Gaussian}
\end{equation}
assuming that the mean square  displacement 
\begin{equation}
\left \langle \Delta r^2(t) \right \rangle=
  \left \langle  \frac{1}{N} \sum_{i=1}^N \big (\mathbf r_i(t)-\mathbf r_i(0) \big )^2 \right \rangle
\label{Eq:msd}
\end{equation}
obeys Gaussian statistics. To test this
hypothesis we first compute the mean square displacement 
$\left \langle \Delta r^2(t) \right \rangle$.
Fig.~\ref{fig:msdeps08} shows the resulting $\left \langle \Delta r^2(t) \right \rangle$
for $\varepsilon=0.8$ and several volume
fractions. One clearly observes a
ballistic regime for small times with a crossover to diffusive
behavior around $t\sim 1$. 
%with diffusive behavior for long times due to particle number conservation. 
%

\begin{figure}
\includegraphics[width=0.47\textwidth]{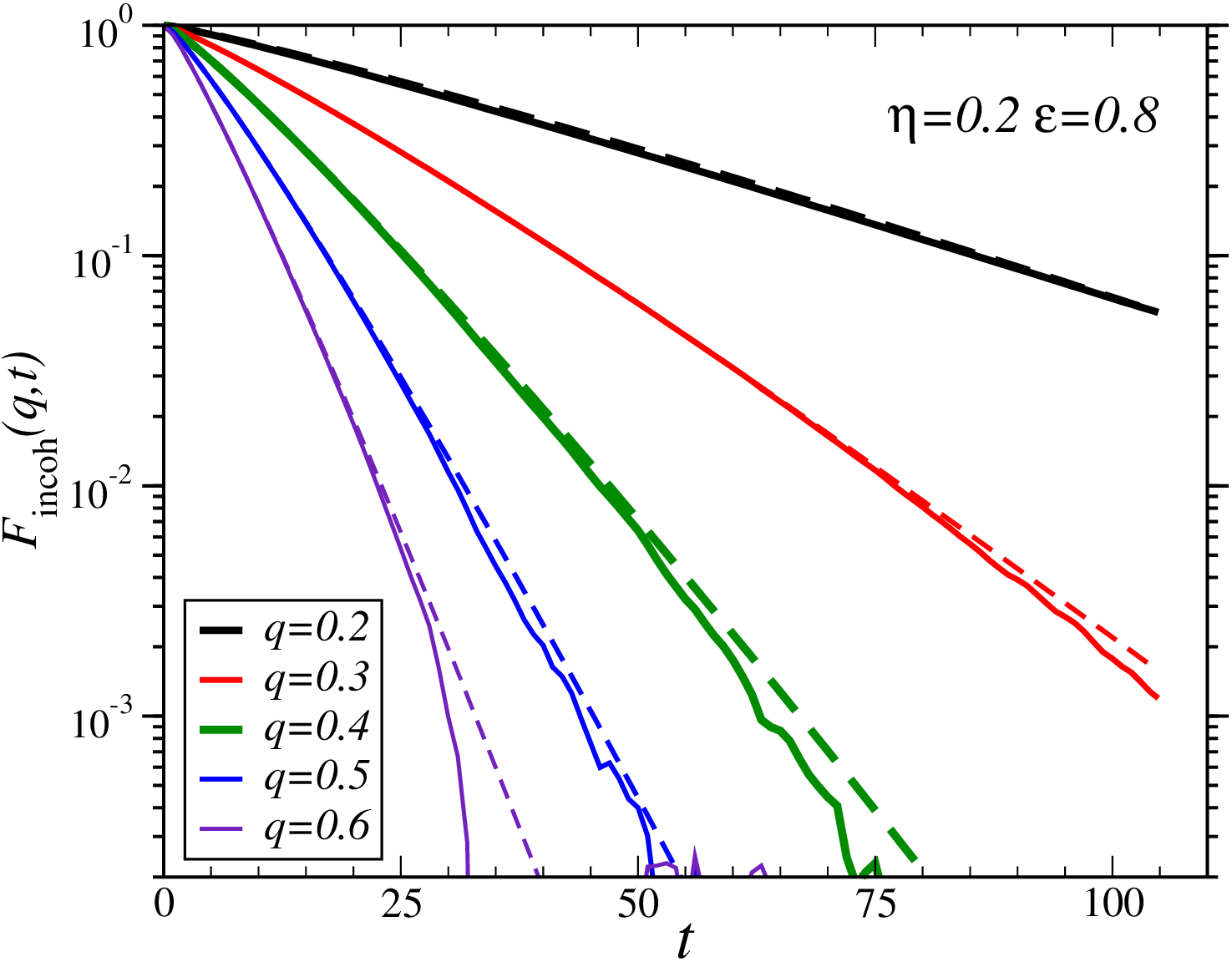}
%\includegraphics[width=0.47\textwidth]{fig.dir/fsmsd_short_eta02_eps08_v3.eps}
%%no longer:
%%\includegraphics[width=0.47\textwidth]{fig.dir/fsmsd_short_eta02_eps09_v2.eps}
\caption{Intermediate incoherent scattering function $F_\mathrm{incoh}(\mathbf q,t)$
using Eq.~(\ref{eq:fqtdef}) (solid lines) and for comparison the 
Gaussian approximation using Eqs.~(\ref{Eq:Gaussian}) and (\ref{Eq:msd}) (dashed lines.)
}
\label{fig:Gaussian}
\end{figure}

The computed 
$\left \langle \Delta r^2(t) \right \rangle$ is then substituted in
the Gaussian approximation of Eq.~(\ref{Eq:Gaussian}) and compared to
the full scattering function in Fig.~\ref{fig:Gaussian}. The Gaussian
approximation works very well for the densities under consideration,
in particular for small $q$. %, where $F_\mathrm{incoh}(\mathbf q,t)=e^{\frac{-q^2D t}{6}}$ 
%is given by the hydrodynamic expression. 

% for decision of which see notes p.141

\begin{figure}
\includegraphics[width=0.47\textwidth]{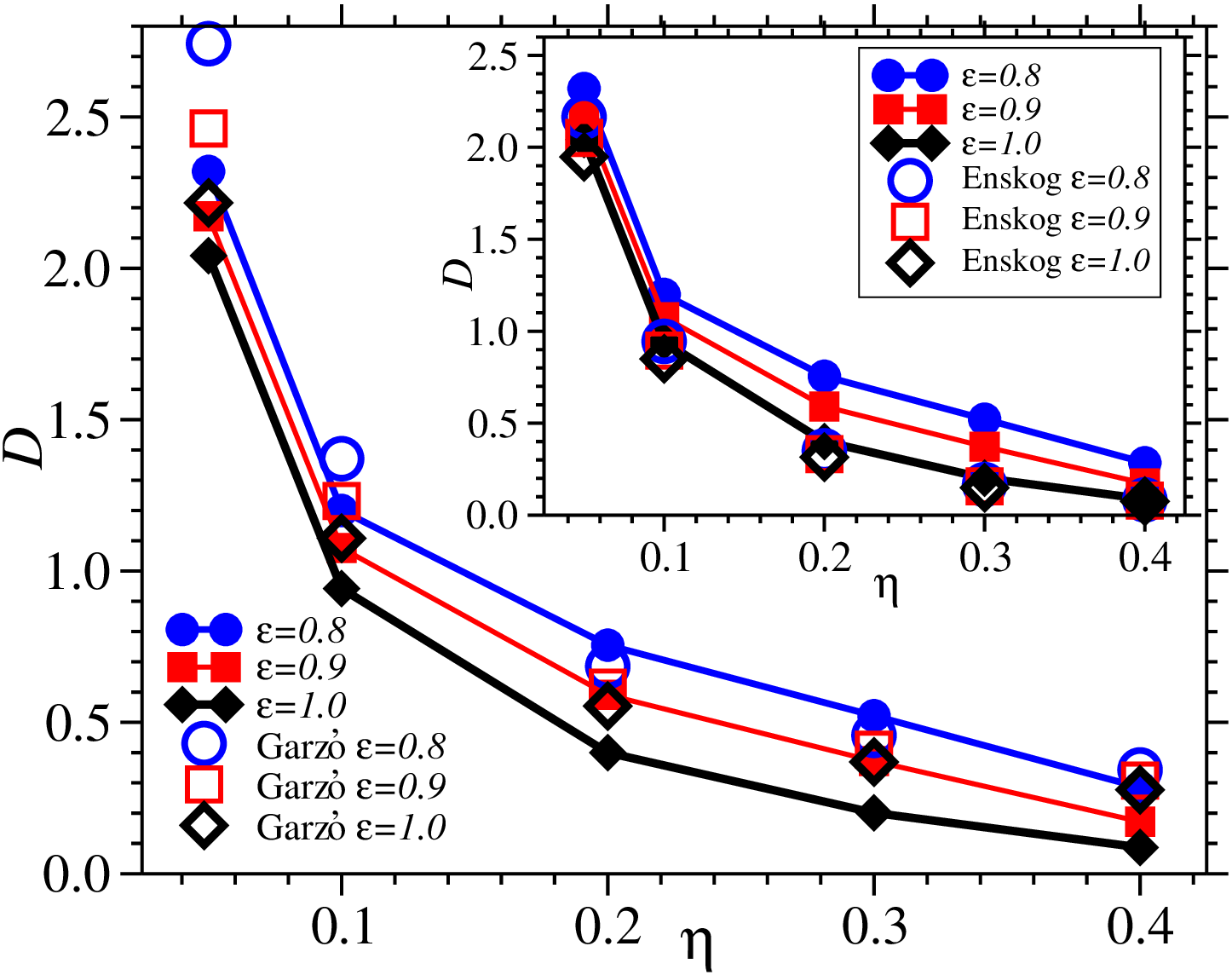}
\caption{Self diffusion constant $D$ as a function of volume fraction $\eta$.
  The filled symbols with lines for the eye are obtained via linear fits to 
  the mean-square displacement for large times. Garz\'{o} results are
corresponding to Eq.~(2.10) of \cite{Garzo2007} and the Enskog result corresponds 
%to Eq.~(14.20) of \cite{poeschel}. 
to Eq.~(5) of \cite{fiege2009}. 
%The dashed lines are $D=D_0 (\eta_{\rm c}-\eta)^{\gamma}$ where 
%$D_0=D(\eta=0.4)/\left ((\eta_{\rm c}-0.4)^{\gamma}\right )$ and 
%$\gamma=2.58$ and $\eta_{\rm c}=0.64056/0.716281/0.869646$ for $\varepsilon=1.0/0.9/0.8$
%as obtained in the fits for $\tau(\eta)$.
}
\label{fig:Dofeta}
\end{figure}

We can also extract the self diffusion,
\begin{equation}
D =  \lim_{t \to \infty}\frac{\left \langle \Delta r^2(t) \right \rangle}{6 t},
\end{equation}
via a linear fit to $\left \langle \Delta r^2(t) \right \rangle$ at 
long times.
The resulting $D$
is plotted in Fig.~\ref{fig:Dofeta} as a function of density (filled symbols)
and compared with theoretical predictions (open symbols).
As expected, the diffusion constant decreases strongly with density.
Whereas the prediction of Enskog (see Eq.~(5) of \cite{fiege2009}) is 
in excellent agreement for the elastic case (see inset), the prediction 
of Garz{\'{o}} \cite{Garzo2007} is very good for the inelastic case and 
$\eta > 0.1$.

\begin{figure}
\includegraphics[width=0.47\textwidth]{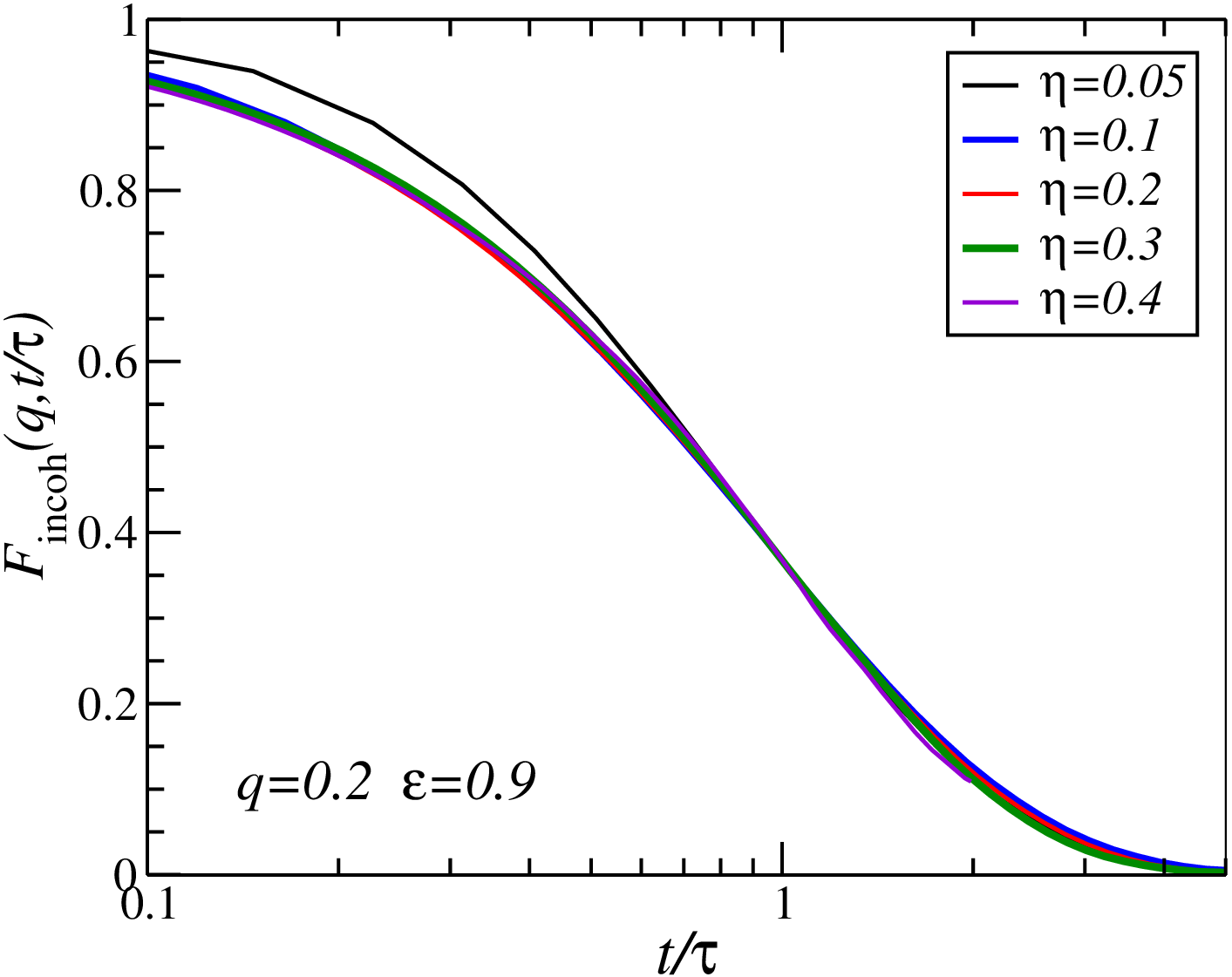}
\caption{Time-density superposition for the incoherent scattering
  function.}
\label{fig:superpos}
\end{figure}

For the glass transition in elastic systems, one observes dynamic scaling as
the transition is approached. In other words, the scattering function
does not depend separately on time and control parameter --- either temperature
or density ---  but only on the ratio $t/\tau$. We have
tested this time-density superposition principle by plotting 
$F_\mathrm{incoh}(\mathbf q,t/\tau)$ for five volume fractions in Fig.~\ref{fig:superpos}. Even
though the volume fractions under consideration are far away from the
critical value, the data collapse for $\eta \ge 0.1$.  %$t/\tau\gtrsim 1$.

%\pagebreak
 
\section{IV Intermediate Coherent Scattering Function and Longitudinal Current  Correlation}
\label{sec:Fqt}

\begin{figure}
\includegraphics[width=0.47\textwidth]{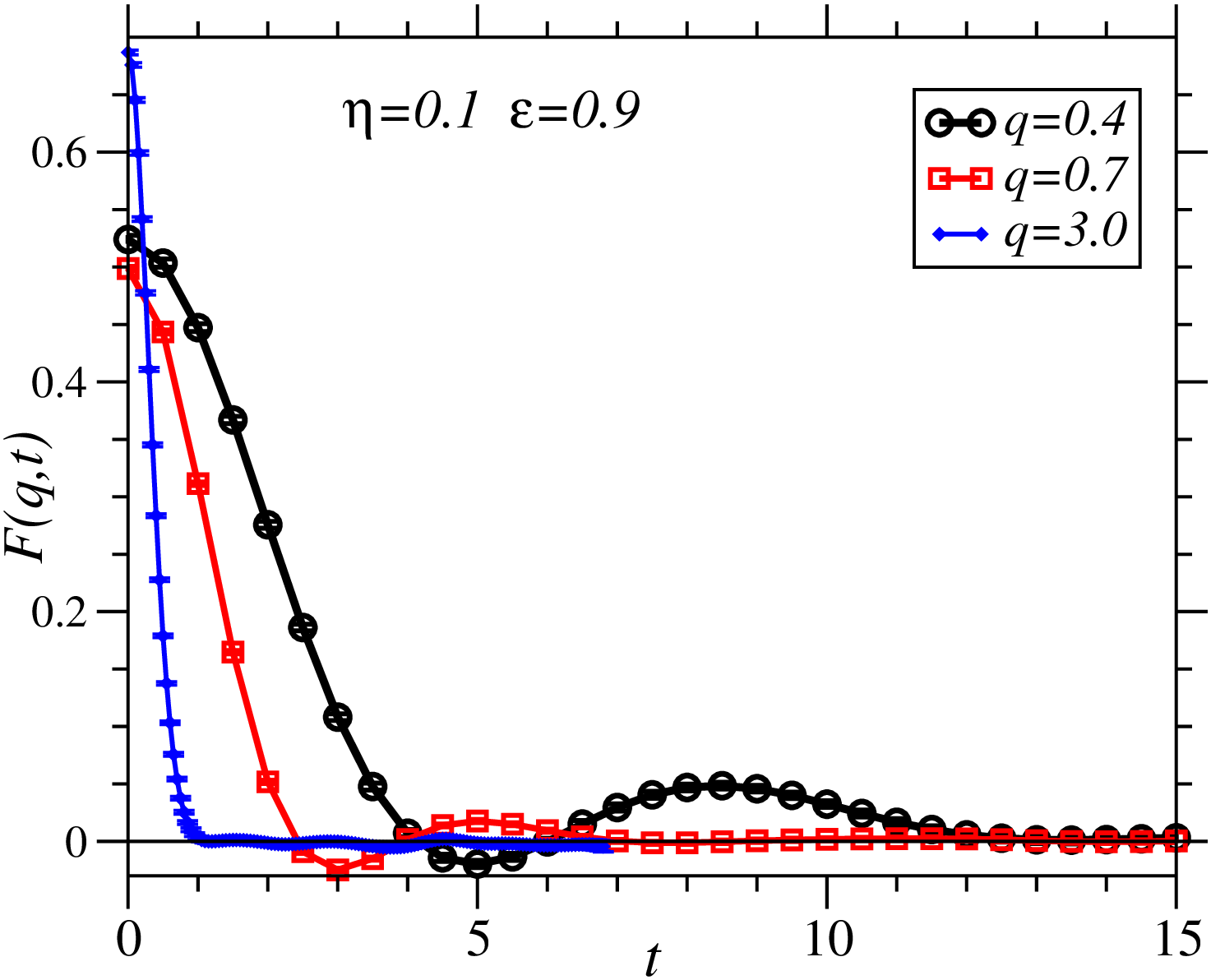}
\caption{Intermediate coherent scattering function $F(q,t)$.}
\label{fig:Fqt}
\end{figure}

% S(q,omega) dynamic structure factor (Hansen McD. p.219 
% F(q,t) intermediate scattering function or coherent interm.sc.fct.
Information about the collective dynamics and in particular
collective density fluctuations is contained in the intermediate
{\it coherent} scattering function, 
defined by
\begin{equation}
F(\mathbf q,t)= \left \langle \frac{1}{N}\sum_{i,j=1}^N 
e^{i\mathbf q \cdot (\mathbf r_i(t)-\mathbf r_j(0)} \right \rangle.
\end{equation}
In the hydrodynamic regime, i.e. small wave numbers, we expect to see
sound modes. This expectation is indeed born out by the data with an
example shown in Fig.~\ref{fig:Fqt} for volume fraction $\eta=0.1$, restitution 
coefficient $\epsilon=0.9$, and for
several $q$ values. We observe oscillations which are overdamped
for large $q$.

\begin{figure}
\includegraphics[width=0.47\textwidth]{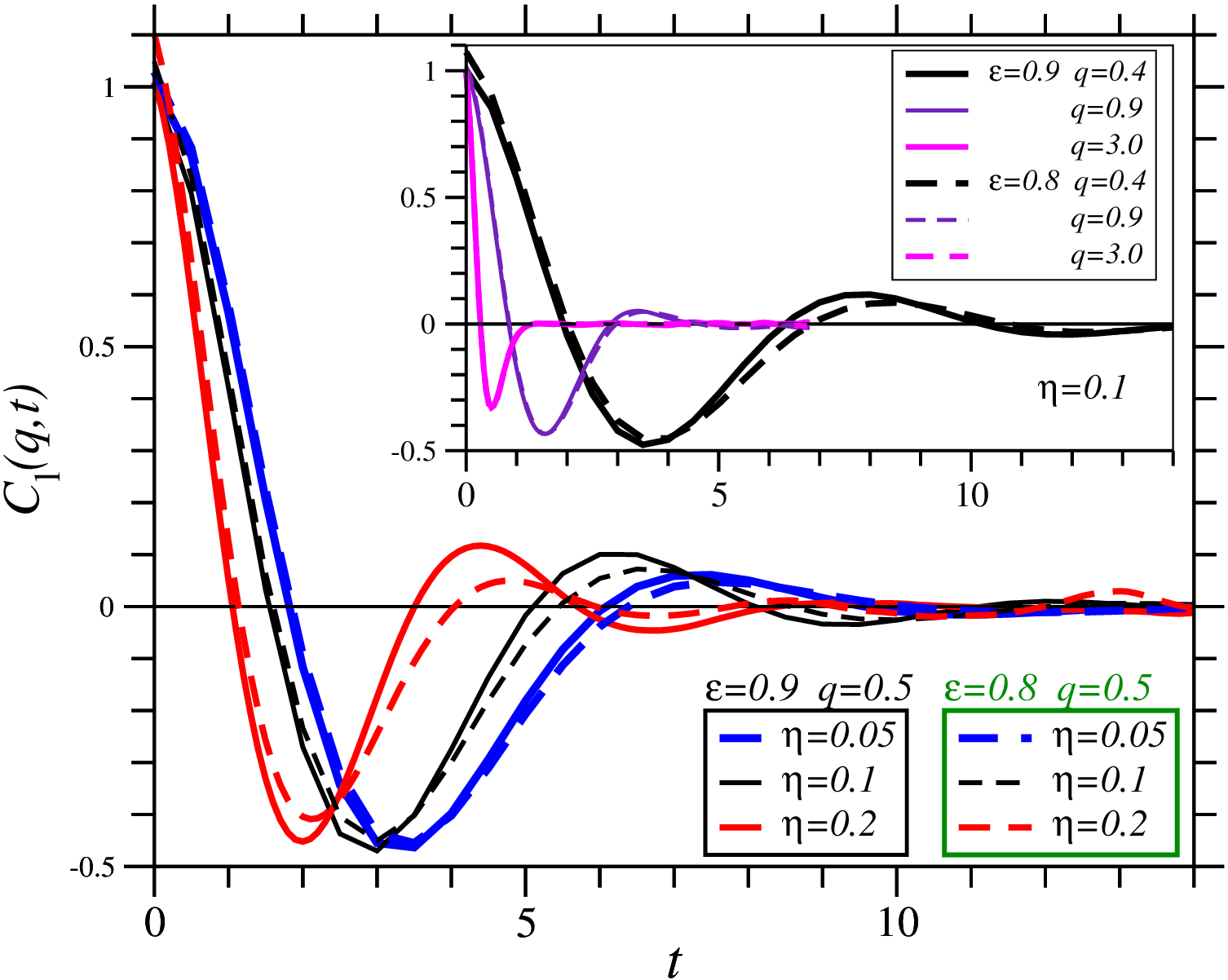}
\caption{Longitudinal current correlation as a function of time for
  several densities; inset: variation with $q$.
}
\label{fig:Clt}
\end{figure}
\vspace{1cm}

A detailed analysis of the coherent correlation in terms of damped
sound waves and temperature fluctuations will be given in Sec.\
VI, using a hydrodynamic model discussed in the next section. Here we
consider the longitudinal current correlation to obtain an
approximation to the sound velocity which we analyze in dependence on
volume fraction and inelasticity. The
correlation of the longitudinal current is defined as
\begin{eqnarray}
C_l(\mathbf q,t)&=&\left \langle \frac{1}{N}\sum_{i,j=1}^N \frac{1}{q^2} 
  \left (\mathbf q\cdot \mathbf v_i \right )  \, 
  \left (\mathbf q \cdot \mathbf v_j \right )
e^{i\mathbf q \cdot (\mathbf r_i(t)-\mathbf r_j(0)}\right \rangle \nonumber \\
&=& -\frac{1}{q^2}
\partial_t^2F(\mathbf q,t).
\end{eqnarray}
In Fig.~\ref{fig:Clt} we show data for two values of restitution and volume
fractions $0.05 \leq \eta \leq 0.2$. For all parameters we observe
well-defined oscillations which are more strongly damped for the more
inelastic system.

\begin{figure}
\includegraphics[width=0.47\textwidth]{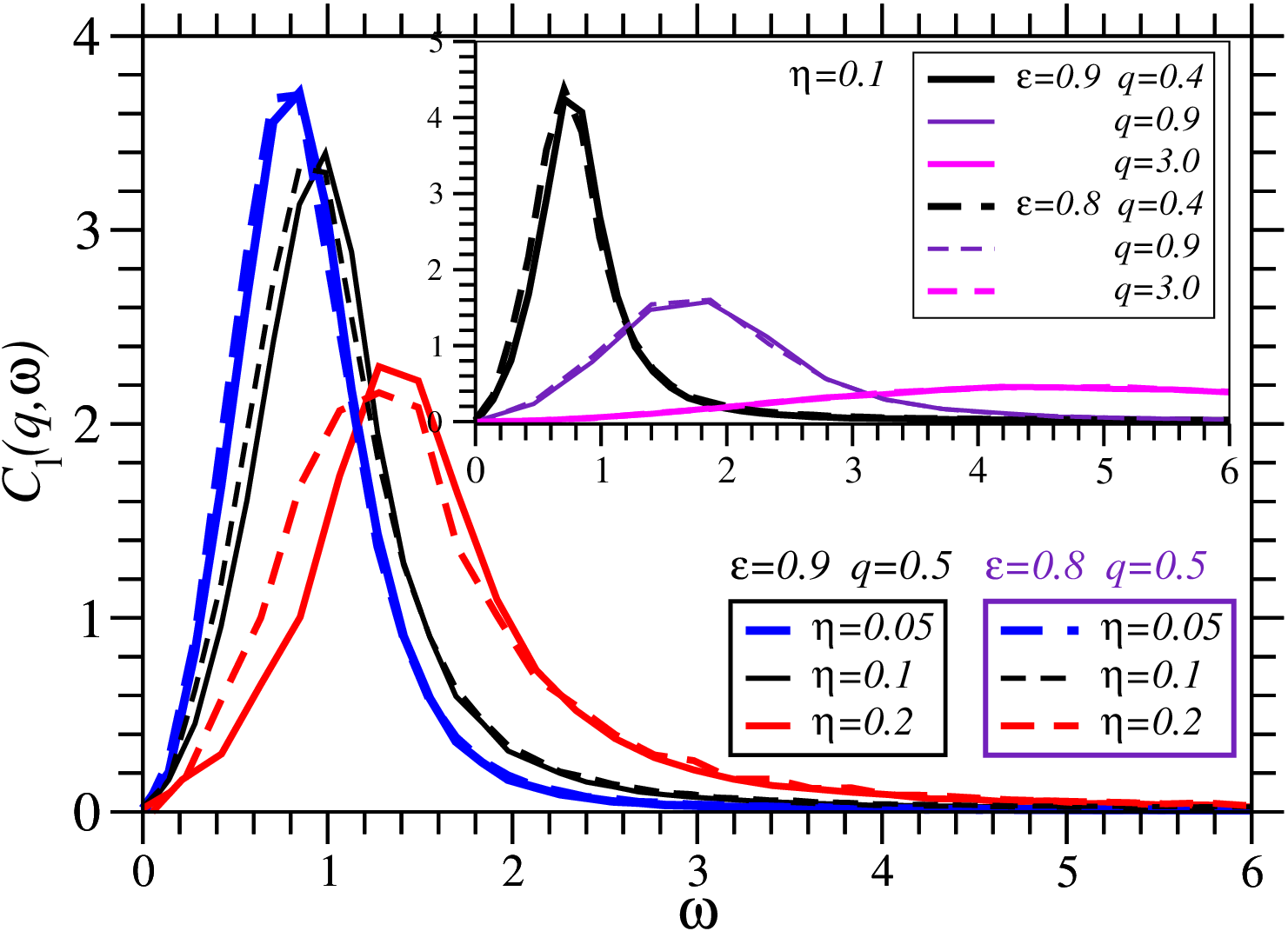}
\caption{Longitudinal current correlation as a function of angular frequency.}
\label{fig:Clomega}
\end{figure}

 In Fig.~\ref{fig:Clomega} we plot $C_l(\mathbf q,\omega)$
the corresponding Fourier transform of the current correlation.

%and determine peak position and half width at half maximum. 

\begin{figure}
\includegraphics[width=0.47\textwidth]{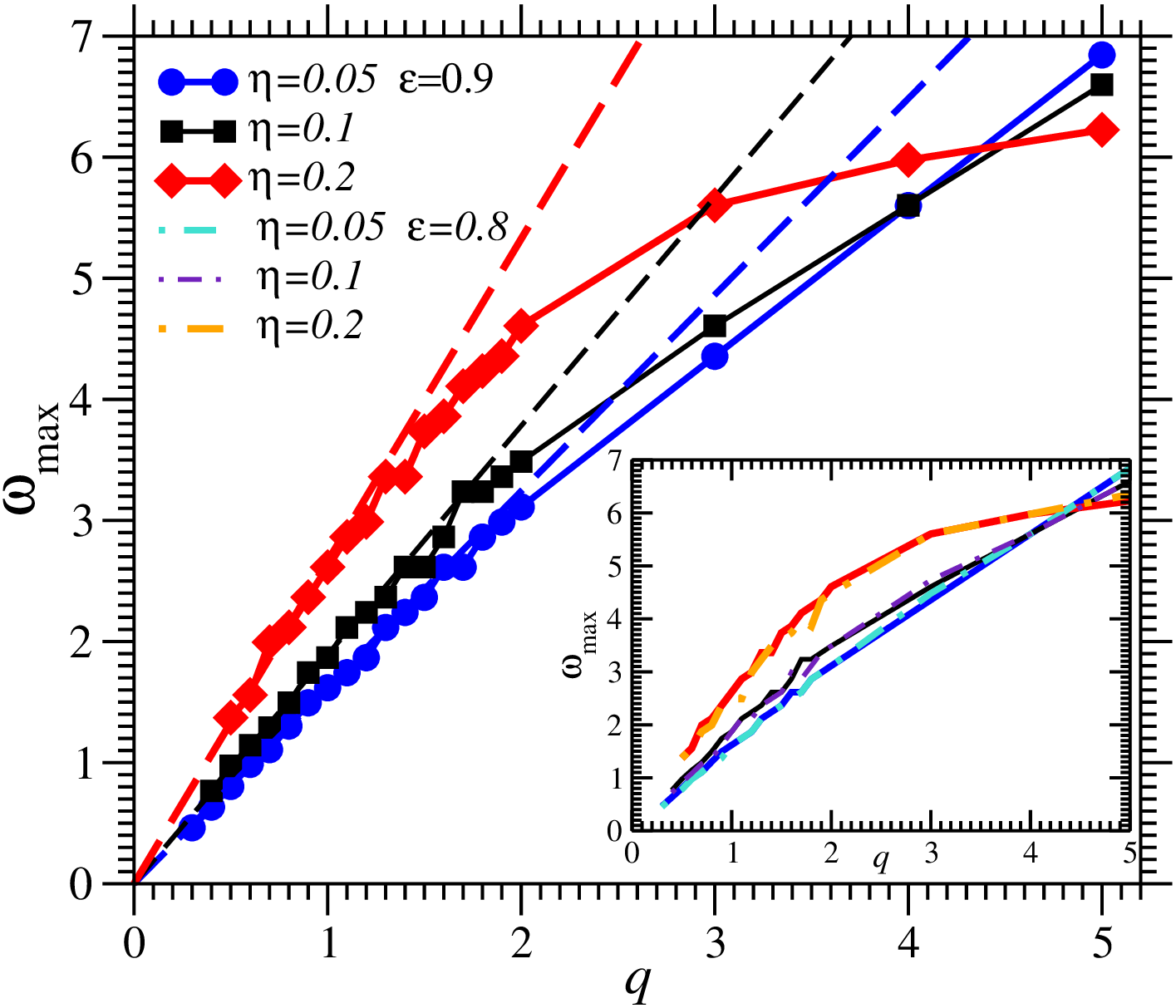}
\caption{Position of the maximum, $\omega_\mathrm{max}$ of longitudinal current
  correlation $C_l(\mathbf q,\omega)$. 
%Slopes $c$ of the line fits are
%for $\varepsilon=0.9$ and
%for $\eta=0.05/0.1/0.2$ $c=1.62\pm0.01/1.89\pm0.01/2.66\pm0.04$  and (not shown) 
%for $\varepsilon=0.8$ $c=1.58\pm0.02/1.81\pm0.02/2.57\pm0.04$.}
Solid lines and filled symbols are for $\epsilon=0.9$ and dashed lines 
indicate linear fits with slopes as listed in Table~\ref{table:cvalues}. The inset 
shows for comparison $\omega_\mathrm{max}$ for $\epsilon=0.9$ (solid lines) 
and for $\epsilon=0.8$ (dash-dotted lines).}
\label{fig:omegamax}
\end{figure}

%The speed of sound can be estimated from the position of the maximum
The position of the maximum of $C_l(\mathbf q,\omega)$ can be used to 
estimate the speed of sound. The peak position 
$\omega_\mathrm{max}$  as a function of wave number $q$
is shown in Fig.~\ref{fig:omegamax}.
As shown in 
the inset, the peak position does not depend on $\varepsilon$. For small wave numbers a
linear dispersion is observed (dashed lines in  Fig.~\ref{fig:omegamax}), 
while deviations from linear behavior for
larger wave numbers are more pronounced for the denser systems. 

%In the following we compare in Sec.\ VI %\ref{sec:Fqomega} 
%quantitatively these simulation results 
%with hydrodynamic predictions as described  in Sec.\ V. %\ref{sec:hydrodynamics}.

%The half width at half maximum is shown in Fig.\ref{width} as a
%function of wave number.
%\begin{figure}
%\includegraphics[width=0.47\textwidth]{fig.dir/HWofq_etas_v3.eps}
%\caption{Half width at half maximum of longitudinal current
%  correlation $C_l(\mathbf q,f)$}
%\label{width}
%\end{figure}

\section{V Fluctuating Hydrodynamics}
\label{sec:hydrodynamics}

%{\bf Annette: Eq. for $\mathbf u$ hat auch driving and dissip.,
%ist das hier so okay? Auch: Ich hab' probiert dimension so weit 
%moeglich allgemein mit $d$ zu lassen.}\\
%%
In this section we compute $S(q,\omega)$ from fluctuating
hydrodynamics. Our presentation follows closely the work of van Noije
et al. \cite{noije1999}, except that we take care to conserve momentum at
each instant of time, thereby avoiding a divergence of the static
structure factor.

The hydrodynamic equations for the number density $n$ and flow velocity
$\mathbf u$ are the same
as for an elastic fluid. However the equation for the temperature
differs due to the energy dissipation in collisions and the energy
input due to driving:
\begin{equation}
%\partial_t T=D_T\Delta T-\frac{2p}{3n_0}\nabla \mathbf u -\Gamma +m\xi_0^2 +\theta
\partial_t T=D_T\Delta T-\frac{2p}{d n}\nabla \mathbf u -\Gamma +m\xi_0^2 +\theta.
\end{equation}
Here we present results in $d=3$ dimensions. 
The energy dissipation due to collisions, $\Gamma$, is
estimated as $\Gamma= 2 T \nu_\mathrm{coll} \frac{1-\varepsilon^2}{2 d}$ 
with the collision frequency $\nu_\mathrm{coll}$. The input of
kinetic energy due to driving is given by $m\xi_0^2$, $p$ denotes the
pressure and $D_T$ the thermal diffusivity. We have ignored nonlinear
terms involving the flow field because we will consider only linear
hydrodynamics. In the stationary state the energy dissipation in
collisions and the energy input due to driving balance on average:
\begin{equation}
\Gamma_0=m\xi_0^2.
\end{equation}
We expand in fluctuations around the stationary state: $n=n_0+\delta
n, T=T_0+\delta T$ and $\Gamma=\Gamma_0+\delta \Gamma$.
The collision frequency should be proportional to 
the density, the pair correlation function at contact, $\chi$, and 
the thermal velocity: $\nu_\mathrm{coll}\propto n \chi T^{1/2}$, 
hence linearization around
the stationary state $\Gamma_0$ yields: 
$\Gamma\sim \Gamma_0(1+\frac{\delta n}{n_0}
     +\frac{1}{\chi}\frac{\mathrm {d}\chi}{\mathrm {d}n}\delta n
     +\frac{3\delta T}{2 T_0}
)$.

Following van Noije et al. \cite{noije1999}, we consider a hydrodynamic
description of a granular fluid based on conservation of particle
number and momentum and the relaxation of temperature to its
stationary value, $T_0$. The transverse momentum decouples so that we are
left with three equations for the fluctuating density $\delta n$,
the longitudinal flow velocity $u(\mathbf q,t)=\mathbf q \cdot \mathbf
u/q$, and the fluctuating temperature $\delta T$:
\begin{eqnarray}
\partial_t \delta n(\mathbf q,t)&=&- i q n_0 u(\mathbf q,t)\\
\partial_t u(\mathbf q,t)&=&
  -\frac{iq}{mn_0}\left (\frac{\partial p}{\partial n}\delta n(\mathbf q,t)
  +\frac{\partial p}{\partial T}\delta T(\mathbf q,t)\right )\\
  &-&\nu_\mathrm{l} q^2 u(\mathbf q,t) +\xi_\mathrm{l}(\mathbf q,t)\nonumber\\
\partial_t \delta T(\mathbf q,t)&=&-D_T q^2 \delta T(\mathbf q,t) -i q  \frac{2p_0}{d n_0} u(\mathbf q,t)\\
 &-&\Gamma_0 \left (\frac{\delta n(\mathbf q,t)}{n_0}
     +\frac{1}{\chi}\frac{\mathrm {d}\chi}{\mathrm {d}n}\delta n(\mathbf q,t)
     +\frac{3}{2}\frac{\delta T(\mathbf q,t)}{T_0} \right ) \nonumber \\
 &+& \theta(\mathbf q,t),\nonumber
\end{eqnarray}
where $D_T=\frac{2 \kappa}{d n_0}$ with the heat conductivity $\kappa$, and 
where $\nu_\mathrm{l}$ is the longitudinal viscosity. Fluctuating
hydrodynamics for an elastic
fluid ($\epsilon=1$) is based on internal noise,
$\xi^{\mathrm{in}}_\mathrm{l}$ and $\theta^{\mathrm {in}}$, consistent with the
fluctuations-dissipation theorem. Here we consider a randomly driven
system: the particles are kicked randomly, giving rise to external
noise in the equation for the velocity as well as the temperature. 
The external contributions are 
\begin{equation}
\xi_{\mathrm{l}}^{\mathrm {ex}}(\mathbf r,t)
   =\frac{1}{n_0}\sum_i \xi_{i \mathrm l}(t) \delta(\mathbf r-\mathbf r_i)
\end{equation}
and
\begin{equation}
\theta^{\mathrm {ex}}(\mathbf r,t)=\frac{2 m}{d n_0}\sum_i 
   \mathbf v_i \cdot \bm\xi_i(t) \delta(\mathbf r-\mathbf r_i(t)),
\end{equation}
%Noije (15)
with variance 
\begin{equation}
\langle\xi_{\mathrm{l}}^{\mathrm {ex}}(\mathbf q,t)\xi_{\mathrm{l}}^{\mathrm {ex}}
   (-\mathbf q,t^\prime)\rangle=
  V    \frac{\xi_0^2}{n_0}\delta (t-t^\prime) (1-\delta_{\mathbf q,\mathbf 0})
\end{equation}
and
\begin{equation}
\langle \theta^{\mathrm {ex}}(\mathbf q,t)\theta^{\mathrm {ex}}(-\mathbf q,t^\prime)\rangle
  = V \frac{4 m T_0}{d n_0} \xi_0^2 \delta (t-t^\prime) (1-\delta_{\mathbf q,\mathbf 0}).
\end{equation}
Here we have taken care of global momentum conservation, as realized
by our driving mechanisms involving pairs of particles. These
terms occur only for $q=0$ and ensure that the driving force vanishes
at zero wave number.
%and which are of the order of $1/N$.
%{\bf Annette: Ist das auch fuer $\xi$ richtig so?, auch habe ich $V$ dazufuegt  
%       wie Noije (19),(20). Wuerde auch dazu passen, dass ich in Timos Notizen 
%       Glg.(7),(8) 1/V nicht erwarte, was dann in Glg.(9) das V geben wuerde.}
Including both types of noise, 
$\xi_{\mathrm{l}}=\xi_{\mathrm{l}}^{\mathrm
  {in}}+\xi_{\mathrm{l}}^{\mathrm {ex}}$ and
$\theta_{\mathrm{l}}=\theta_{\mathrm{l}}^{\mathrm
  {in}}+\theta_{\mathrm{l}}^{\mathrm {ex}}$,
 as suggested by Noije et al. \cite{noije1999}, one
obtains 
%Noije (19),(20)
\begin{equation}
 \langle\xi_{\mathrm{l}}(\mathbf q,t)\xi_{\mathrm{l}} (-\mathbf q,t^\prime)\rangle=
    V \left (\frac{\xi_0^2}{n_0} + \frac{2 \nu_{\mathrm l} T_0 q^2}{m n_0} \right )
         \delta (t-t^\prime) (1-\delta_{\mathbf q,\mathbf 0})
\end{equation}
and 
\begin{equation}
\langle \theta(\mathbf q,t)\theta(-\mathbf q,t^\prime)\rangle
  =4 V \left ( \frac{m T_0 \xi_0^2}{d n_0}  + \frac{2 \kappa T_0^2 q^2}{d^2 n_0^2} \right )
       \delta (t-t^\prime) (1-\delta_{\mathbf q,\mathbf 0}).
\end{equation}

To complement the above equations, we need an expression for the
pressure $p$ in terms of the density and temperature. Since the driven
granular gas is far from equilibrium, we cannot expect that a
thermodynamic description including an equation of state should hold
in general. Nevertheless for small to moderate inelasticities an
equation of state has been found empirically (Eq.~(17.29) in \cite{poeschel}):
% Brilliantov (17.29)
$ p\approxeq n T \left (1+2\eta \chi (1+\varepsilon)\right )$. We use the 
Carnahan-Starling approximation (for $d=3$)
\begin{equation}
\chi = \frac{\left (1 - \eta/2 \right )}{\left (1-\eta \right )^3}.
\end{equation}
This leaves
us with two unknown parameters in the hydrodynamic description, namely
the longitudinal viscosity, $\nu_{\mathrm l}$, and the thermal diffusivity, $D_T$.

 The linearized equations can be solved for the frequency and
 wave number dependent correlation functions, $S(q,\omega)$ and $
 C_l(q,\omega)=\frac{\omega^2}{q^2} S(q,\omega)$. Of particular interest is the
 pole structure in the complex $\omega$-plane, describing damped sound
 modes and the decay of temperature fluctuations. The latter can be
 either diffusive or with a finite relaxation rate, depending on wave number $q$. For
 $D_T q^2 \ll \frac{3\Gamma_0}{2T_0}$, the thermal diffusivity can be
 ignored and we have poles at
\begin{eqnarray}
\omega_T&=&\pm i \frac{3\Gamma_0}{2T_0}\\
\omega_s&=&\pm cq \pm i \gamma q^2.  
\end{eqnarray}
The latter correspond to sound modes with sound velocity
\begin{equation}
c^2=v_{\mathrm{th}}^2-\frac{2 p_0}{3 m n_0} 
                \left ( 1 + \frac{n_0}{\chi} \frac{\partial \chi}{\partial n} \right )
%notice: below is without (n/chi dchi/dn)   term
%=\frac{T}{3m}\big(1+8(1+\varepsilon ) \eta g(a)\big)
\label{c1}
\end{equation}
where $v_{\mathrm{th}}^2=\frac{1}{m} \left (\frac{\partial p}{\partial n} \right )_T$, and damping
\begin{equation}
2\gamma= \nu_{\mathrm l} + \frac{4 p_0 T_0}{3 \Gamma_0 m n_0} 
    \left ( \frac{1}{3} \left [ 1 + \frac{n_0}{\chi} \frac{\partial \chi}{\partial n} \right ]
             + \frac{p_0}{d T_0 n_0} \right ).
% below ?
%2\gamma=\frac{3\Gamma_0}{2T_0}+\frac{2T_0}{3\Gamma_0}\frac{2p}{3\rho_0} \big(1+(1+\varepsilon)\eta g(a)\big)
\end{equation}

In the opposite limit $D_T q^2 \gg \frac{3\Gamma_0}{2T_0}$, we recover
ordinary hydrodynamics of an elastic fluid. The sound speed is given
by the adiabatic value
\begin{equation}
c^2=v_s^2=v_{\mathrm {th}}^2+\frac{2 p_0^2}{d m T_0 n_0^2}
\label{c2}
\end{equation} and the temperature decay is diffusive 
$\omega_T=\pm i D_T q^2\frac{v_{\mathrm {th}}^2}{v_s^2}$. 
%$\omega_T=\pm i D_T q^2\frac{v_s^2}{v_{th}^2}$. 

In general, we expect to see a crossover, when 
 $D_T q_{\mathrm c}^2 = \frac{3\Gamma_0}{2T_0}$. In order to estimate $q_{\mathrm c}$, we
 use the Enskog values for the collision frequency in three dimensions 
and the thermal diffusivity: 
\begin{eqnarray}
%notes p.127, Brilliantov S.199 Eq.~(20.30)
\nu_\mathrm{coll}&=& \omega_{\mathrm E} = 4\pi  \chi n_0a^2\sqrt{\frac{T_0}{\pi m}} \\
D_T&=&\frac{75}{d(1+\varepsilon)(49-33\varepsilon)n_0 a^2 \chi}\sqrt{\frac{T_0}{\pi
    m}}.
\end{eqnarray}
These yield for the crossover wave number
\begin{equation}
q_{\mathrm c}^2 a^2=\frac{6(1-\varepsilon^2)(1+\varepsilon)(49-33\varepsilon)\chi^2 36\eta^2}{75\pi}.
\label{eq:qc}
\end{equation}
Numerical values of estimated $q_{\mathrm c} a$ for the simulated volume fractions $\eta$ and 
inelasticities $\epsilon$ are given in Table~\ref{table:qcvalues}

\begin{table}[htb]
\centering% NICHT \begin{center}
\begin{tabular}{|c|c|c|} \hline \hline
  $\eta$ & $\epsilon$ & $q_{\mathrm c} a$ \\ \hline \hline
  0.05 &   0.8 &  0.21  \\ \hline
  0.05 &   0.9 &  0.14  \\ \hline
  0.1  &   0.8 &  0.48  \\ \hline
  0.1  &   0.9 &  0.33  \\ \hline
  0.2  &   0.8 &  1.29  \\ \hline
  0.2  &   0.9 &  0.89  \\ \hline
\end{tabular}
%\begin{tabular}{ccc} 
%  $\eta$ & $\epsilon$ & $q_{\mathrm c} a$ \\ \hline \hline
%  0.05 &   0.8 &  0.21  \\ \hline
%  0.05 &   0.9 &  0.14  \\ \hline
%  0.1  &   0.8 &  0.48  \\ \hline
%  0.1  &   0.9 &  0.33  \\ \hline
%  0.2  &   0.8 &  1.29  \\ \hline
%  0.2  &   0.9 &  0.89  \\ \hline
%\end{tabular}
\caption{Estimates for $q_{\mathrm c} a$ using Eq.~(\ref{eq:qc}).}
\label{table:qcvalues}
\end{table}

For the case of $D_T q^2 \approx \frac{3\Gamma_0}{2T_0}$ it is necessary to use the 
more general solution  for the dynamic structure factor
%\onecolumngrid
\begin{widetext} 
\begin{equation}
S(q,\omega)= n_0 q^2  \left (
   \frac{\left [ \omega^2 + \left ( 3 \gamma_0 \omega_{\mathrm E}+ D_T q^2 \right )^2 \right ]
         \left [ \frac{\xi_0^2}{n_0} + \frac{2 \nu_{\mathrm l} T_0 q^2}{m n_0} \right ]
         + q^2 \left (\frac{p_0}{m n_0 T_0} \right )^2 
         \left [ \frac{4 m T_0 \xi_0^2}{d n_0} + \frac{4 D_T T_0^2 q^2}{d n_0} \right ] 
    }
    {\left | {\mathrm {det}} M \right |^2}
    \right ),
\label{S8fitnumerator}
\end{equation}

where we have used $\nu_\mathrm{coll}=\omega_{\mathrm E}$, the abbreviation
$\gamma_0=\frac{1-\varepsilon^2}{2 d}$ and where \\
\begin{eqnarray}
\left | {\mathrm {det}} M \right |^2 & = &
  \left [ - \omega^2 
            \left ( 3 \gamma_0 \omega_{\mathrm E} + D_T q^2 + \nu_{\mathrm l} q^2 \right )
          + q^2 \left ( 3 \gamma_0 \omega_{\mathrm E} v_{\mathrm {th}}^2 
                      - \frac{2 p_0 \gamma_0 \omega_{\mathrm E}}{m n_0}
                       \left \{ 1 + \frac{n_0}{\chi}\frac{\partial \chi}{\partial n} \right \}
                      + v_{\mathrm {th}}^2 D_T q^2 
                \right )
  \right ]^2 \nonumber \\
 & + & \left [ \omega^3 - \omega q^2 
                  \left ( 3 \nu_{\mathrm l} \gamma_0 \omega_{\mathrm E} 
                         + \nu_{\mathrm l} D_T q^2
                         + v_{\mathrm {th}}^2
                         + \frac{2 p_0^2}{d m T_0  n_0^2}
                  \right )
       \right ]^2.
\label{S8fitdenominator}
\end{eqnarray}
\end{widetext} 
%%%%%%%%%%%%%%%%%%%%%%%%%%%%%%%%%%%%%%%%%%%%%%%%%%%%%%%%%
\section{VI Coherent Scattering Function and Transport 
Coefficients}
\label{sec:Fqomega}

\begin{figure}
\includegraphics[width=0.47\textwidth]{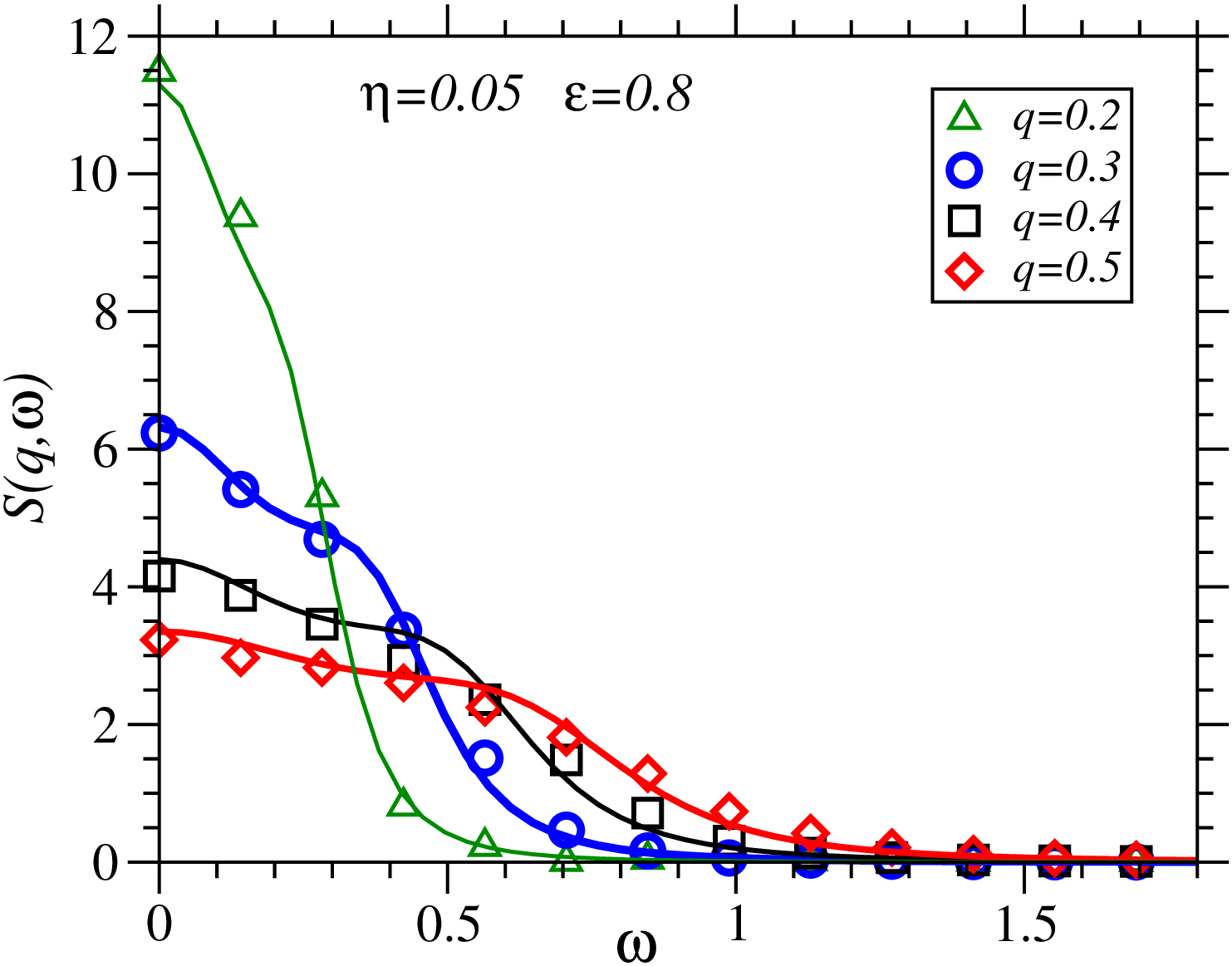}
\caption{Dynamic structure factor $S(q,\omega)$ for $\eta=0.05,\, \varepsilon=0.8$ 
  and $q=0.2$ -- $0.5$. Symbols indicate simulation results obtained via Fourier 
transform of $F(q,t)$ and lines indicate fits with Eq.~(\ref{S8fitnumerator}). }
\label{Sofom_eta005eps08q03-05}
\end{figure}

\begin{figure}
\includegraphics[width=0.47\textwidth]{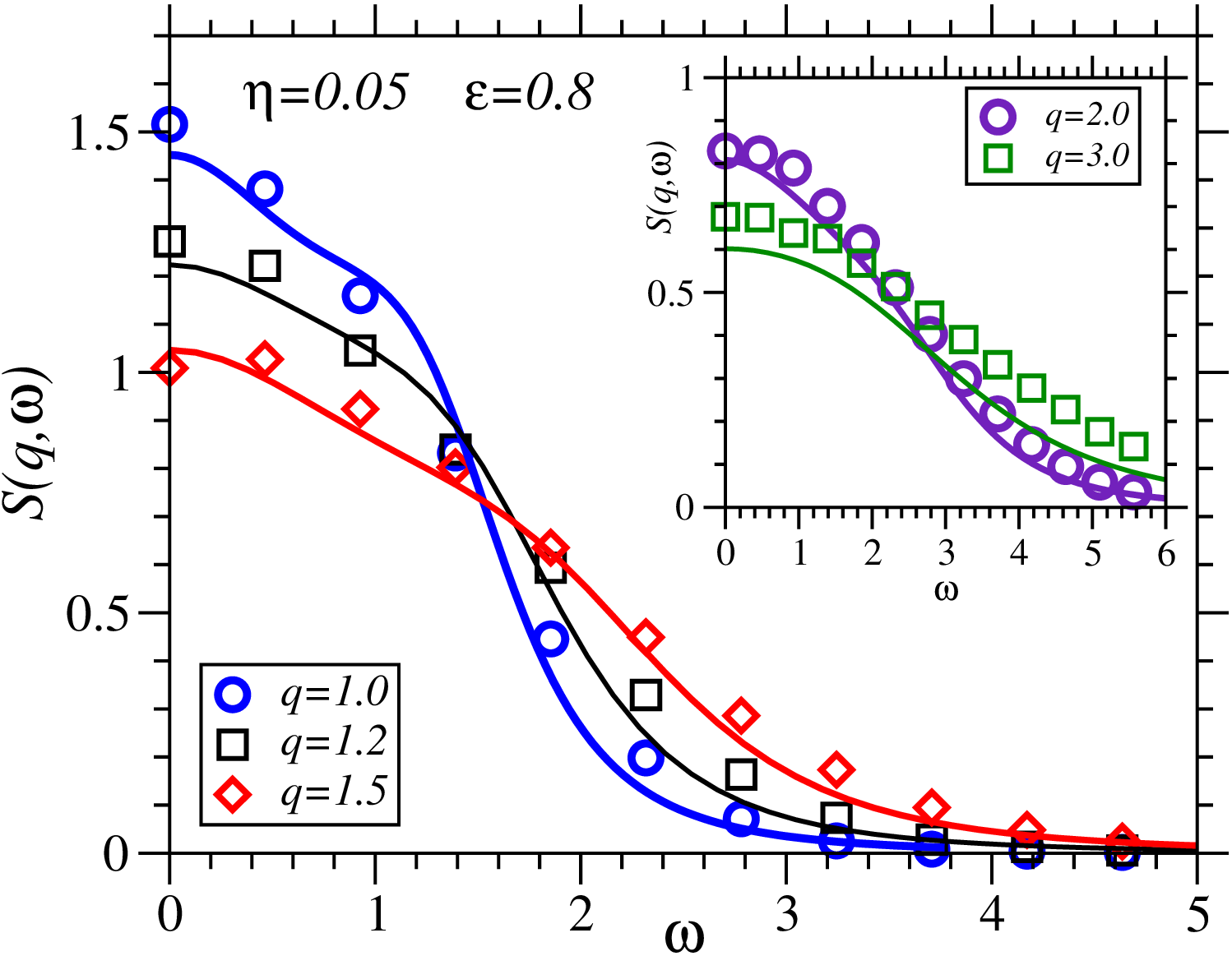}
\caption{$S(q,\omega)$ for $\eta=0.05,\, \varepsilon=0.8$ and $q=1.0$ -- $3.0$. }
\label{Sofom_eta005eps08q1-3}
\end{figure}

\begin{figure}
\includegraphics[width=0.47\textwidth]{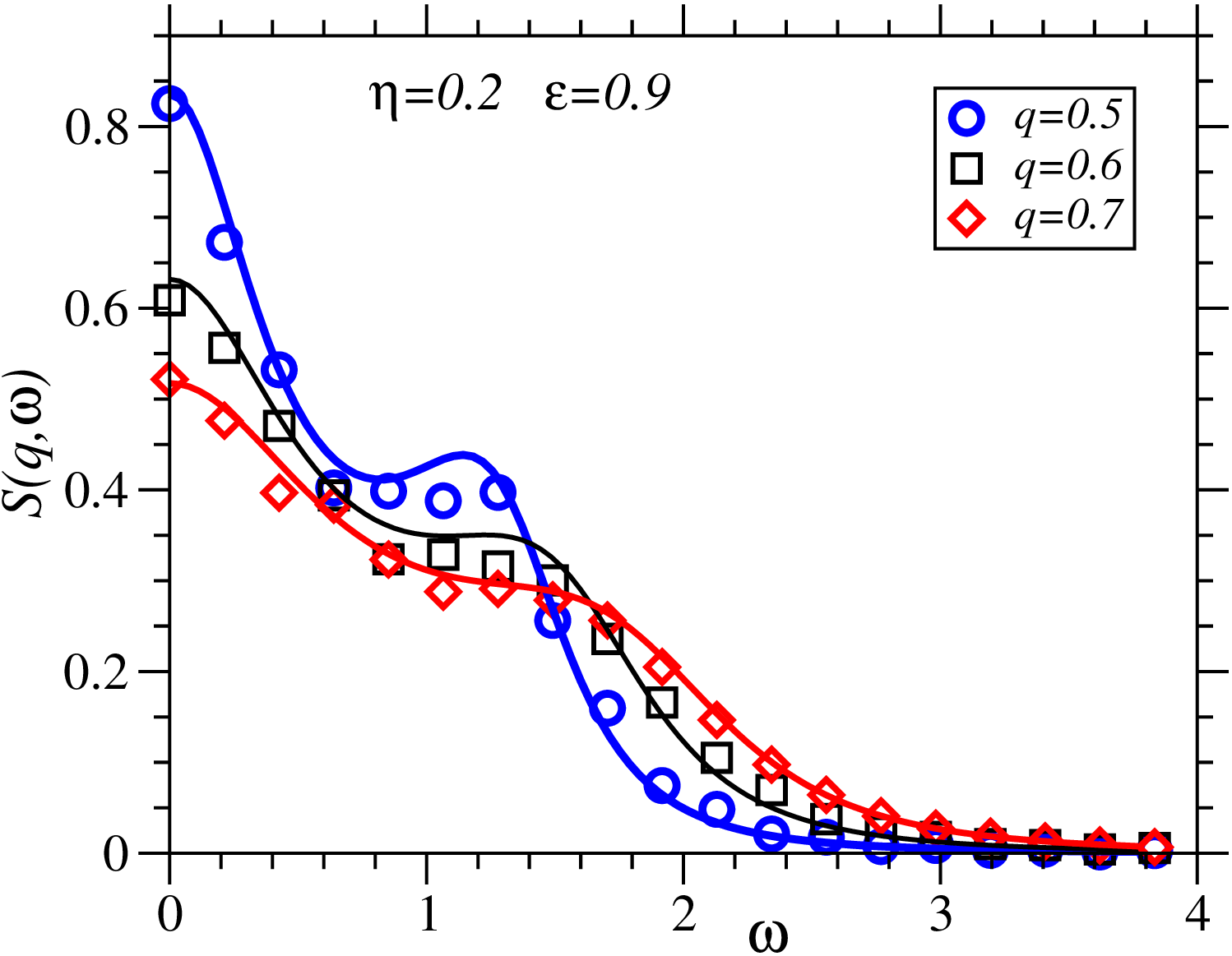}
\caption{$S(q,\omega)$ for $\eta=0.2,\, \varepsilon=0.9$ and $q=0.5$ -- $0.7$. }
\label{Sofom_eta02eps09q05-07}
\end{figure}

\begin{figure}
\includegraphics[width=0.47\textwidth]{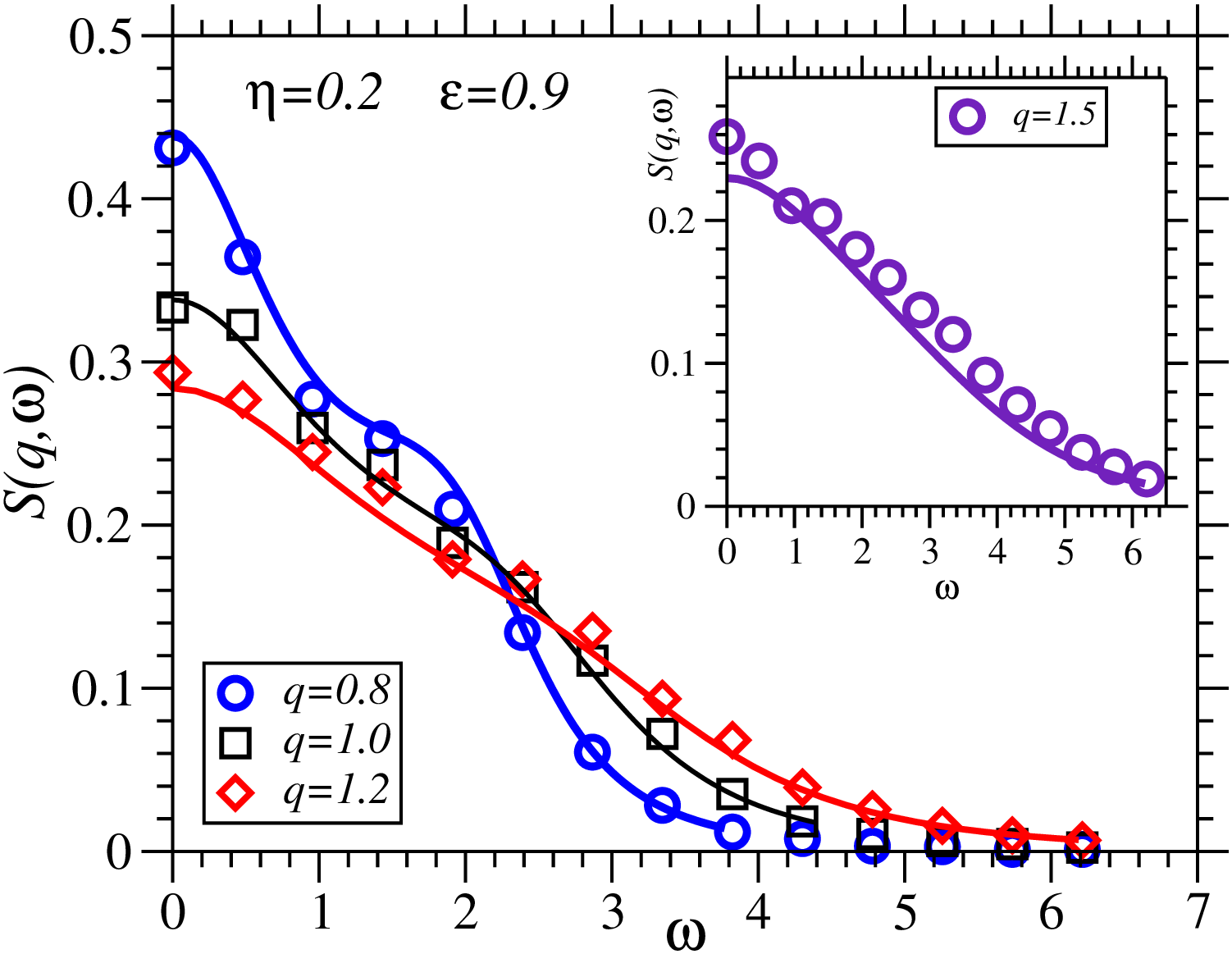}
\caption{$S(q,\omega)$ for $\eta=0.2,\, \varepsilon=0.9$ and $q=0.8$ -- $1.5 $.}
\label{Sofom_eta02eps09q08-16}
\end{figure}

According to the $q_\mathrm{c}$ estimates given in  Table~\ref{table:qcvalues}, 
we see that our data are neither clearly in the hydrodynamic regime
nor in the inelastic regime, but in general the two relaxation terms in
the equation for the temperature are comparable in magnitude. Hence, we
fit the simulation results of the dynamic structure factor 
to the full expression for $S(q,\omega)$ as given in 
Eqs.~(\ref{S8fitnumerator}) and (\ref{S8fitdenominator}). We allow
for two fit parameters, $D_T$ and $\nu_l$, with all other parameters
determined by the approximate equation of state.
The best fits (solid line) are shown in
Figs.~\ref{Sofom_eta005eps08q03-05} -- \ref{Sofom_eta02eps09q08-16};
in comparison with the simulation data (symbols) for 
$S(q,\omega)$.
We find excellent agreement not only for very small $q$, for which we would expect 
best agreement with the hydrodynamic equations,  but also for $ q \lesssim 1.0$.
Both features, the shoulder due to the sound wave as well as the damping, are 
quantitatively in agreement with Eqs.~(\ref{S8fitnumerator}) and (\ref{S8fitdenominator}).
Similarly, we find very good agreement for the $\eta=0.1$ results.

%\begin{figure}
%\includegraphics[width=0.47\textwidth]{fig.dir/Sofom_eta005eps08q06-08.eps}
%\caption{$S(q,\omega)$ for $\eta=0.05 \varepsilon=0.8$ and $q=0.6$ -- $0.8$. }
%\label{Sofom_eta005eps08q06-08}
%\end{figure}

\begin{figure}
\includegraphics[width=0.47\textwidth]{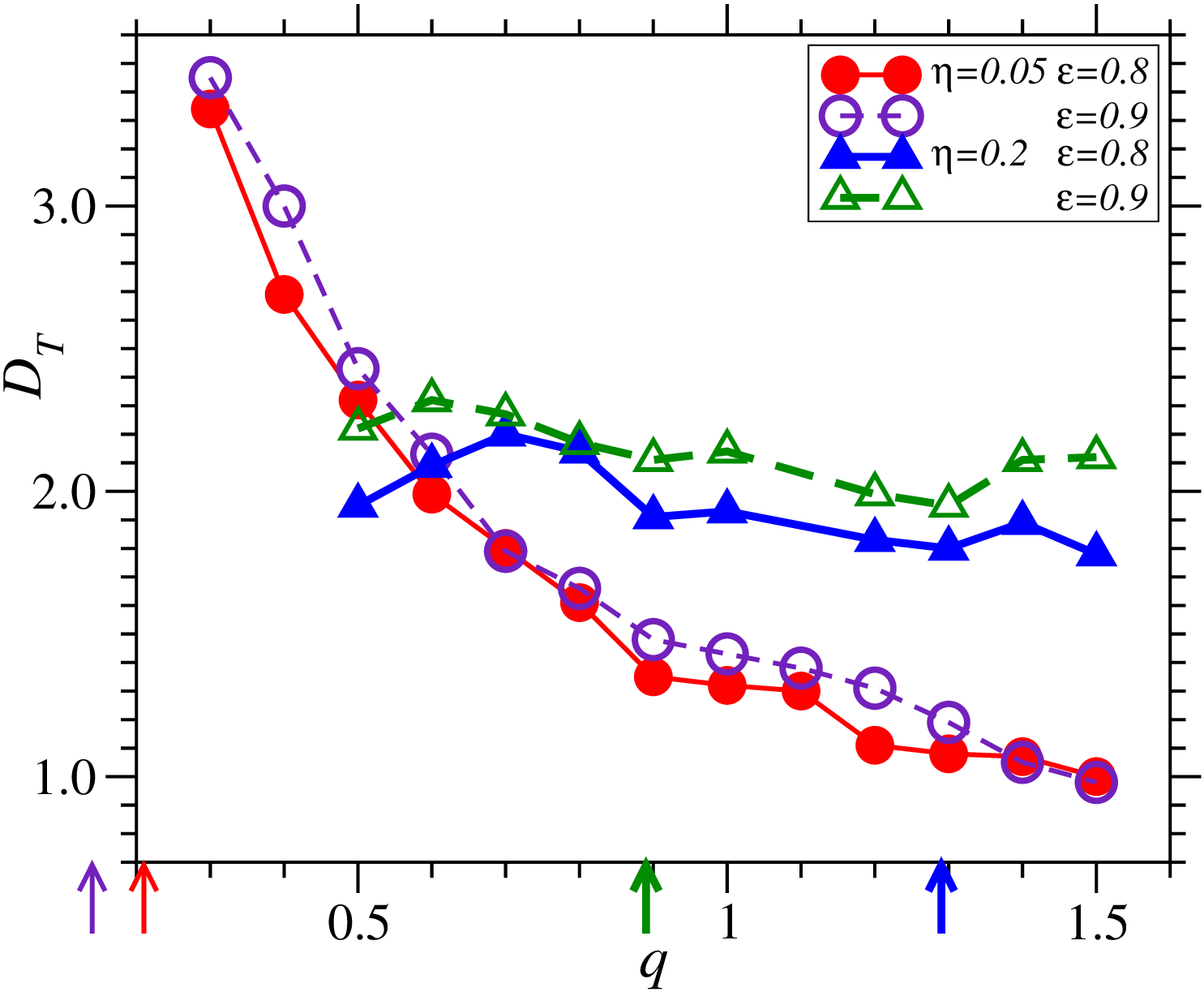}
\caption{Thermal diffusivity $D_T(q)$ obtained via fits to $S(q,\omega)$ with 
  Eq.~(\ref{S8fitnumerator}). The arrows indicate the $q_c$ values
  from Table~\ref{table:qcvalues}.}
\label{fig:DTofq}
\end{figure}

\begin{figure}
\includegraphics[width=0.47\textwidth]{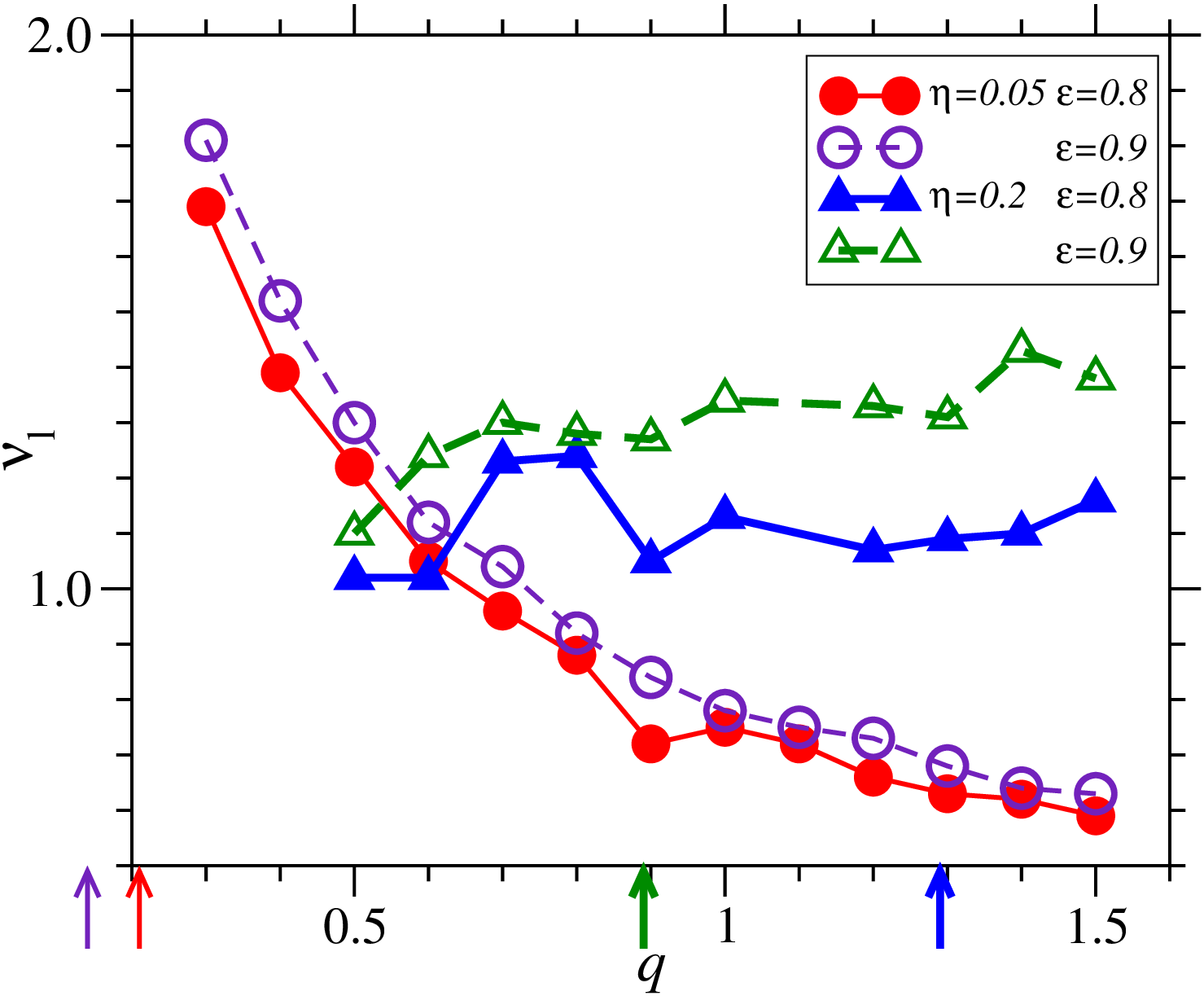}
\caption{Longitudinal viscosity $\nu_{\rm l}(q)$ obtained via fits to $S(q,\omega)$ with 
  Eq.~(\ref{S8fitnumerator}). 
  The arrows indicate the $q_c$ values from Table~\ref{table:qcvalues}.}
\label{fig:nulofq}
\end{figure}

The corresponding best fit parameters  are the transport coefficients 
$D_T$ and $\nu_\mathrm{l}$  which are shown 
graphically in Figs.~\ref{fig:DTofq} \& \ref{fig:nulofq}. The fits require
$q$-dependent transport coefficients because we consider wave numbers
outside the hydrodynamic regime. It is difficult to estimate the
hydrodynamic regime, but we need at least  $q<q_c$ (see Table~\ref{DTnultable}),
corresponding to $D_T q^2 < \frac{3\Gamma_0}{2T_0}$. For
$\eta=0.2$, we are able to reach this regime  and indeed  find that 
$D_T$ and $\nu_\mathrm{l}$ are approximately independent of $q$.
For $\eta=0.05$ even the smallest $q$ values are not in the
hydrodynamic regime yet, and for $\eta=0.1$ the smallest wave numbers
are in the crossover regime. As far as temperature fluctuations are
concerned, the diffusive regime is restricted to larger wave numbers
$D_T q^2 > \frac{3\Gamma_0}{2T_0}$, so that $D_T$ can only be
extracted from an intermediate range of $q$-values, such that $q>q_c$
but $q$ still small enough to ignore higher order terms in $q$. Again,
for $\eta=0.2$ this seems possible, whereas for $\eta=0.05$ our data
are not sufficient.

% as expected, we find, on the other hand,
%for $\eta=0.05$ a strong $q$ 
%dependence even for the smallest investigated $q$. 
Tabulated in Table~\ref{DTnultable} is a quantitative comparison 
of the fit results for small $q$ with the  theoretical predictions 
%Noije Appendix
for $D_T=\frac{2 \kappa}{d n}$ and 
$\nu_\mathrm{l}=\frac{1}{\rho} \left ( 
            \frac{2 \eta_{\mathrm{shear}} (d-1)}{d} + \zeta \right )$, where 
$\eta_{\mathrm{shear}}$ and $\zeta$ are shear and bulk viscosity respectively.
For the comparison with Brilliantov et al. we use Eqs.~(20.13) \& (20.30) 
of Ref.~\cite{poeschel} for 
$\eta_{\mathrm{shear}}$ and $\kappa$ respectively and $\zeta$ of Eq.~(32) of
Ref.~\cite{Dufty1997}.
For the predictions of Dufty et al. we used Eqs.~(29), (30) and (32) 
of Ref.~\cite{Dufty1997}
and for the predictions of Garz{\'{o}} et al. we used Eqs.~(B1), (2.2), (3.8), (2.3) 
and (3.9) of Ref.~\cite{Garzo2007}.
%#Dufty eta al Physica A 240,212(1997)  Eq.~(32)
%        $bulkvis=(1.0+$epsvalue)*(2.0/9.0)*($n0**2)*$g2*sqrt($pivalue*$m*$T0)
%  Dufty et al.: shear visc.(29)     therm.cond.(30)   bulk visc.(32)  \cite{Dufty1997} 
%
%Garzo et al.: bulkvisc (B1) correction a2 otherwise same as Dufty \cite{Garzo2007}
%              shearvisc (2.2) & (3.8)
%              therm.cond. (2.3) & (3.9)

%S(q) first peak at about 2*pi/d approx 6 -> so q > 1 not nonsense accord. to S(q)
% DT(q^2) except for eta=0.1 does not reach plateau for small q

\begin{table}[htb]
\centering% NICHT \begin{center}
\begin{tabular}{|r||c|c||c|c|} \hline \hline
\multicolumn{5}{|c|}{$\eta=0.05$ } \\ \hline 
 & \multicolumn{2}{|c|} {$\varepsilon=0.8$} & \multicolumn{2}{|c|} {$\varepsilon=0.9$} \\ \hline 
  & $D_T$ & $\nu_{\rm l}$ & $D_T$ & $\nu_{\rm l}$ \\ \hline 
Fit Results: $q=0.2$ & 4.72  &  2.55 & 4.63 & 3.23  \\ \hline
  $q=0.3$ &   3.34 &  1.69  & 3.45 & 1.81 \\ \hline
  $q=0.4$ &   2.69 &  1.39  & 3.00 & 1.52 \\ \hline
Brilliantov et al. \cite{poeschel}& 3.19 & 2.26 & 3.54 & 2.25 \\ \hline
Dufty et al. \cite{Dufty1997} & 4.71 & 2.82  & 4.07 & 2.77 \\ \hline
Garz\'{o} et al. \cite{Garzo2007} & 5.62 & 2.78 & 5.06 & 2.67 \\ \hline \hline \hline
\multicolumn{5}{|c|}{$\eta=0.1$ } \\ \hline 
 & \multicolumn{2}{|c|} {$\varepsilon=0.8$} & \multicolumn{2}{|c|} {$\varepsilon=0.9$} \\ \hline 
  & $D_T$ & $\nu_{\rm l}$ & $D_T$ & $\nu_{\rm l}$ \\ \hline 
Fit Results:  $q=0.3$ &   2.23 &  1.20  & 2.67 & 1.70 \\ \hline
  $q=0.4$ &   2.25 &  1.02  & 2.42 & 1.25 \\ \hline
  $q=0.5$ &   2.15 &  1.07  & 2.33 & 1.20 \\ \hline
Brilliantov et al. \cite{poeschel}& 1.39 & 1.13 & 1.55 & 1.13 \\ \hline
Dufty et al. \cite{Dufty1997} & 2.67 & 1.69  & 2.42 & 1.71 \\ \hline
Garz\'{o} et al. \cite{Garzo2007} & 2.81 & 1.53 & 2.53 & 1.48 \\ \hline \hline \hline
\multicolumn{5}{|c|}{$\eta=0.2$ } \\ \hline 
 & \multicolumn{2}{|c|} {$\varepsilon=0.8$} & \multicolumn{2}{|c|} {$\varepsilon=0.9$} \\ \hline 
  & $D_T$ & $\nu_{\rm l}$ & $D_T$ & $\nu_{\rm l}$ \\ \hline 
Fit Results:  $q=0.5$ &   1.95 &  1.02  & 2.22 & 1.10 \\ \hline
  $q=0.6$ &   2.09 &  1.02  & 2.32 & 1.24 \\ \hline
  $q=0.7$ &   2.20 &  1.23  & 2.27 & 1.30 \\ \hline
Brilliantov et al. \cite{poeschel}& 0.52 & 0.83 & 0.57 & 0.85 \\ \hline
Dufty et al. \cite{Dufty1997} & 2.03 & 1.63  & 2.01 & 1.72 \\ \hline
Garz\'{o} et al. \cite{Garzo2007} & 1.40 & 1.15 & 1.26 & 1.15 \\ \hline \hline \hline
\end{tabular}
\caption{Comparison of theoretical predictions and fit results for $D_T$ and $\nu_\mathrm{l}$.}
\label{DTnultable}
\end{table}

We find best agreement between the simulation results for the smallest $q$ and 
the predictions of Dufty et al. \cite{Dufty1997} and fairly good agreement
with the predictions of Garz\'{o} et al. \cite{Garzo2007}.

Finally, we compare the speed of sound as obtained from the maximum of
the current correlation with the predictions from the hydrodynamic
theory in either the inelastic regime (see Eq.~(\ref{c1})) or the diffusive
regime (see Eq.~(\ref{c2})). We find very good agreement 
(see Table~\ref{table:cvalues}) of the 
simulation results with Eq.~(\ref{c2}), implying 
$D_T q^2 \gg \frac{3\Gamma_0}{2T_0}$ and adiabatic sound propagation.
However one should keep in mind that our procedure to extract
the sound velocity from the maximum of the current correlation
yields only an estimate of the sound velocity.

\begin{table}[htb]
\centering% NICHT \begin{center}
\begin{tabular}{|r|c|c|c|} \hline \hline
% & \multicolumn{3}{|c|}{$c$} \\ \cline{2-4}
 & via $\omega_\mathrm{max}(q)$ &
 $D_T q^2 \ll \frac{3\Gamma_0}{2T_0}$  &
 $D_T q^2 \gg \frac{3\Gamma_0}{2T_0}$   \\ \hline 
% & $c$ via $\omega_\mathrm{max}(q)$ & $c$ for 
% $D_T q^2 \ll \frac{3\Gamma_0}{2T_0}$  & $c$ for
% $D_T q^2 \gg \frac{3\Gamma_0}{2T_0}$   \\ \hline \hline
$\epsilon=0.8 \quad \eta = 0.05$ &1.58 & 0.73 & 1.55 \\ \hline
                   $\eta = 0.1$ &1.81 & 0.90 & 1.87 \\ \hline
                   $\eta = 0.2$ &2.57 & 1.37 & 2.79 \\ \hline
$\epsilon=0.9 \quad \eta = 0.05$ &1.62 & 0.74 & 1.56 \\ \hline
                   $\eta = 0.1$ &1.89 & 0.92 & 1.90 \\ \hline
                   $\eta = 0.2$ &2.66 & 1.40 & 2.86 \\ \hline \hline
\end{tabular}
%\begin{tabular}{rccc} 
%\hline \hline
% & via $\omega_\mathrm{max}(q)$ &  
% $D_T q^2 \ll \frac{3\Gamma_0}{2T_0}$  & 
% $D_T q^2 \gg \frac{3\Gamma_0}{2T_0}$   \\ 
%\hline 
%$\epsilon=0.8 \quad \eta = 0.05$ &1.58 & 0.73 & 1.55 \\ \hline
%                   $\eta = 0.1$ &1.81 & 0.90 & 1.87 \\ \hline
%                   $\eta = 0.2$ &2.57 & 1.37 & 2.79 \\ \hline
%$\epsilon=0.9 \quad \eta = 0.05$ &1.62 & 0.74 & 1.56 \\ \hline
%                   $\eta = 0.1$ &1.89 & 0.92 & 1.90 \\ \hline
%                   $\eta = 0.2$ &2.66 & 1.40 & 2.86 \\ \hline \hline
%\end{tabular}
\caption{The speed of sound, $c$, determined via 
the slope of the simulation results for $\omega_\mathrm{max}(q)$ 
(see Fig.~\ref{fig:omegamax}) compared with the predicted values of 
Eq.~(\ref{c1}) in the case of $D_T q^2 \ll \frac{3\Gamma_0}{2T_0}$ 
and with Eq.~(\ref{c2}) in the case of $D_T q^2 \gg \frac{3\Gamma_0}{2T_0}$.}
\label{table:cvalues}
\end{table}

\section{Conclusions and Outlook}

We have investigated a homogeneously driven granular fluid of hard 
spheres at intermediate volume fractions $0.05 \le \eta \le 0.4$ and 
for constant normal restitution coefficients $0.8 \le \varepsilon \le 1.0$. 
Using event-driven simulations we have determined  
time-delayed correlation functions in the stationary state.

We find for the incoherent intermediate scattering function 
that it follows time-density superposition and that it is well 
approximated by the Gaussian 
$F_\mathrm{incoh}(\mathbf q,t)=e^{\frac{-q^2}{6}\left \langle \Delta
    r^2(t) 
  \right \rangle}$, 
where $\left \langle \Delta r^2(t) \right \rangle$ is the mean square
displacement.
The decay time of $F_\mathrm{incoh}(\mathbf q,t)$ 
is rapidly increasing with increasing $\eta$, giving rise to a
corresponding decrease of the diffusion
constant. This 
precursor of a glass transition, which occurs at significantly larger
$\eta$, is  thus present not only in the elastic fluid but 
also in the inelastic case consistent with previous results at larger densities 
\cite{kranz2010,reis2007,reyes2008}.

We also determine the coherent intermediate scattering function
$F(q,t)$, the longitudinal current correlation function 
$C_\mathrm{l}(q,t)$, and their Fourier transforms $S(q,\omega)$,
$C_\mathrm{l}(q,\omega)$. Because we are interested in the long term
dynamics we have simulated comparatively small systems of $N=10000$
particles and averaged over 100 independent simulation runs.
%we are able to determine the coherent intermediate scattering function
%$F(q,t)$, the longitudinal current correlation function 
%$C_\mathrm{l}(q,t)$ and their Fourier transforms $S(q,\omega)$, $C_\mathrm{l}(q,\omega)$.
We observe sound waves in the form of oscillations in $F(q,t)$ and
estimate the sound velocity from the peak of $C_\mathrm{l}(q,t)$.
For a quantitative comparison with the predictions of generalized fluctuating 
hydrodynamics, we use the linear hydrodynamic equations of Noije et al. \cite{noije1999}
and fit the solutions thereof to the simulation results for $S(q,\omega)$.
Depending on wave number and inelasticity the temperature fluctuations are predicted to 
be governed by inelastic collisions or diffusion \cite{noije1999,McNamara1993}.
Our results are consistent with being in the ``standard regime'' \cite{noije1999}
in which the speed of sound is the same as for elastic particles 
(see Table~\ref{table:cvalues}) and 
the damping of the sound wave depends on inelasticity. The most accurate fits 
were obtained assuming generalized hydrodynamic equations which
account for both temperature diffusion as well as dissipation due to
inelastic collisions (Eq.~(\ref{S8fitnumerator})).
The resulting transport coefficients $D$, $D_T$ and $\nu_\mathrm{l}$ 
compare well with the predictions of Dufty et al. \cite{Dufty1997} and 
Garz\'{o} et al. \cite{Garzo2007}.

We conclude that the time delayed correlations of a fluid of
inelastically colliding particles are well described by generalized 
hydrodynamics. It would be interesting to extend our study in several
directions. First, one would like to see still smaller $q$, requiring 
significantly larger systems (and yet also
many independent runs for sufficient statistics). Second, it would be
interesting to go to higher density and study sound propagation as
the glass transition is approached. Finally, time- or
frequency-dependent response functions are largely unexplored.

\begin{acknowledgments}
K.V.L. thanks the Institute of Theoretical Physics, University of G{\"{o}}ttingen,
for financial support and hospitality.
We thank Till Kranz for many interesting discussions.
\end{acknowledgments}

\bibliography{Korrelationen_v8}% Produces the bibliography via BibTeX

\begin{thebibliography}{30}
\expandafter\ifx\csname natexlab\endcsname\relax\def\natexlab#1{#1}\fi
\expandafter\ifx\csname bibnamefont\endcsname\relax
  \def\bibnamefont#1{#1}\fi
\expandafter\ifx\csname bibfnamefont\endcsname\relax
  \def\bibfnamefont#1{#1}\fi
\expandafter\ifx\csname citenamefont\endcsname\relax
  \def\citenamefont#1{#1}\fi
\expandafter\ifx\csname url\endcsname\relax
  \def\url#1{\texttt{#1}}\fi
\expandafter\ifx\csname urlprefix\endcsname\relax\def\urlprefix{URL }\fi
\providecommand{\bibinfo}[2]{#2}
\providecommand{\eprint}[2][]{\url{#2}}

\bibitem[{\citenamefont{Campbell}(1990)}]{campbell90}
\bibinfo{author}{\bibfnamefont{C.~S.} \bibnamefont{Campbell}},
  \bibinfo{journal}{Annu. Rev. Fluid Mech.} \textbf{\bibinfo{volume}{22}},
  \bibinfo{pages}{57} (\bibinfo{year}{1990}).

\bibitem[{\citenamefont{Brey et~al.}(1998)\citenamefont{Brey, Dufty, Kim, and
  Santos}}]{brey98}
\bibinfo{author}{\bibfnamefont{J.~J.} \bibnamefont{Brey}},
  \bibinfo{author}{\bibfnamefont{J.~W.} \bibnamefont{Dufty}},
  \bibinfo{author}{\bibfnamefont{C.~S.} \bibnamefont{Kim}}, \bibnamefont{and}
  \bibinfo{author}{\bibfnamefont{A.}~\bibnamefont{Santos}},
  \bibinfo{journal}{Phys. Rev. E} \textbf{\bibinfo{volume}{58}},
  \bibinfo{pages}{4638} (\bibinfo{year}{1998}).

\bibitem[{\citenamefont{Sela and Goldhirsch}(1998)}]{sela98}
\bibinfo{author}{\bibfnamefont{N.}~\bibnamefont{Sela}} \bibnamefont{and}
  \bibinfo{author}{\bibfnamefont{I.}~\bibnamefont{Goldhirsch}},
  \bibinfo{journal}{J. Fluid Mech.} \textbf{\bibinfo{volume}{361}},
  \bibinfo{pages}{41} (\bibinfo{year}{1998}).

\bibitem[{\citenamefont{Goldhirsch}(2003)}]{goldhirsch03}
\bibinfo{author}{\bibfnamefont{I.}~\bibnamefont{Goldhirsch}},
  \bibinfo{journal}{Ann. Rev. Fluid Mech.} \textbf{\bibinfo{volume}{35}},
  \bibinfo{pages}{267} (\bibinfo{year}{2003}).

\bibitem[{\citenamefont{Tan and Goldhirsch}(1998)}]{tan98}
\bibinfo{author}{\bibfnamefont{M.-L.} \bibnamefont{Tan}} \bibnamefont{and}
  \bibinfo{author}{\bibfnamefont{I.}~\bibnamefont{Goldhirsch}},
  \bibinfo{journal}{Phys. Rev. Lett.} \textbf{\bibinfo{volume}{81}},
  \bibinfo{pages}{3022} (\bibinfo{year}{1998}).

\bibitem[{\citenamefont{Goldhirsch and Zanetti}(1993)}]{goldhirsch93}
\bibinfo{author}{\bibfnamefont{I.}~\bibnamefont{Goldhirsch}} \bibnamefont{and}
  \bibinfo{author}{\bibfnamefont{G.}~\bibnamefont{Zanetti}},
  \bibinfo{journal}{Phys. Rev. Lett.} \textbf{\bibinfo{volume}{70}},
  \bibinfo{pages}{1619} (\bibinfo{year}{1993}).

\bibitem[{\citenamefont{Herbst et~al.}(2004)\citenamefont{Herbst, M{\"{u}}ller,
  Otto, and Zippelius}}]{herbst04}
\bibinfo{author}{\bibfnamefont{O.}~\bibnamefont{Herbst}},
  \bibinfo{author}{\bibfnamefont{P.}~\bibnamefont{M{\"{u}}ller}},
  \bibinfo{author}{\bibfnamefont{M.}~\bibnamefont{Otto}}, \bibnamefont{and}
  \bibinfo{author}{\bibfnamefont{A.}~\bibnamefont{Zippelius}},
  \bibinfo{journal}{Phys. Rev. E} \textbf{\bibinfo{volume}{70}},
  \bibinfo{pages}{051313} (\bibinfo{year}{2004}).

\bibitem[{\citenamefont{Jenkins and Savage}(1983)}]{jenkins83}
\bibinfo{author}{\bibfnamefont{J.~T.} \bibnamefont{Jenkins}} \bibnamefont{and}
  \bibinfo{author}{\bibfnamefont{S.~B.} \bibnamefont{Savage}},
  \bibinfo{journal}{J. Fluid Mech.} \textbf{\bibinfo{volume}{130}},
  \bibinfo{pages}{187} (\bibinfo{year}{1983}).

\bibitem[{\citenamefont{Lun et~al.}(1984)\citenamefont{Lun, Savage, Jeffery,
  and Chepurniy}}]{lun84}
\bibinfo{author}{\bibfnamefont{C.~K.~K.} \bibnamefont{Lun}},
  \bibinfo{author}{\bibfnamefont{S.~B.} \bibnamefont{Savage}},
  \bibinfo{author}{\bibfnamefont{D.~J.} \bibnamefont{Jeffery}},
  \bibnamefont{and}
  \bibinfo{author}{\bibfnamefont{N.}~\bibnamefont{Chepurniy}},
  \bibinfo{journal}{J. Fluid Mech.} \textbf{\bibinfo{volume}{140}},
  \bibinfo{pages}{223} (\bibinfo{year}{1984}).

\bibitem[{\citenamefont{Brilliantov and P{\"{o}}schel}(2004)}]{poeschel}
\bibinfo{author}{\bibfnamefont{N.~V.} \bibnamefont{Brilliantov}}
  \bibnamefont{and}
  \bibinfo{author}{\bibfnamefont{T.}~\bibnamefont{P{\"{o}}schel}},
  \emph{\bibinfo{title}{Kinetic Theory of Granular Gases}}
  (\bibinfo{publisher}{Oxford University Press, Oxford}, \bibinfo{year}{2004}).

\bibitem[{\citenamefont{Lutsko}(2001)}]{lutsko2001}
\bibinfo{author}{\bibfnamefont{J.~F.} \bibnamefont{Lutsko}},
  \bibinfo{journal}{Phys. Rev. Lett.} \textbf{\bibinfo{volume}{86}},
  \bibinfo{pages}{3344} (\bibinfo{year}{2001}).

\bibitem[{\citenamefont{Santos}(2008)}]{santos2008}
\bibinfo{author}{\bibfnamefont{A.}~\bibnamefont{Santos}},
  \bibinfo{journal}{Phys. Rev. Lett.} \textbf{\bibinfo{volume}{100}},
  \bibinfo{pages}{078003} (\bibinfo{year}{2008}).

\bibitem[{\citenamefont{Olafsen and Urbach}(1999)}]{olafsen1999}
\bibinfo{author}{\bibfnamefont{J.~S.} \bibnamefont{Olafsen}} \bibnamefont{and}
  \bibinfo{author}{\bibfnamefont{J.~S.} \bibnamefont{Urbach}},
  \bibinfo{journal}{Phys. Rev. E.} \textbf{\bibinfo{volume}{60}},
  \bibinfo{pages}{R2468} (\bibinfo{year}{1999}).

\bibitem[{\citenamefont{Reyes and Urbach}(2008)}]{reyes2008}
\bibinfo{author}{\bibfnamefont{F.~V.} \bibnamefont{Reyes}} \bibnamefont{and}
  \bibinfo{author}{\bibfnamefont{J.~S.} \bibnamefont{Urbach}},
  \bibinfo{journal}{Phys. Rev. E.} \textbf{\bibinfo{volume}{78}},
  \bibinfo{pages}{051301} (\bibinfo{year}{2008}).

\bibitem[{\citenamefont{Brey and Ruiz-Montero}(2010)}]{brey2010}
\bibinfo{author}{\bibfnamefont{J.~J.} \bibnamefont{Brey}} \bibnamefont{and}
  \bibinfo{author}{\bibfnamefont{M.~J.} \bibnamefont{Ruiz-Montero}},
  \bibinfo{journal}{Phys. Rev. E.} \textbf{\bibinfo{volume}{81}},
  \bibinfo{pages}{021304} (\bibinfo{year}{2010}).

\bibitem[{\citenamefont{Schr{\"{o}}ter
  et~al.}(2005)\citenamefont{Schr{\"{o}}ter, Goldman, and
  Swinney}}]{schroeter05}
\bibinfo{author}{\bibfnamefont{M.}~\bibnamefont{Schr{\"{o}}ter}},
  \bibinfo{author}{\bibfnamefont{D.~I.} \bibnamefont{Goldman}},
  \bibnamefont{and} \bibinfo{author}{\bibfnamefont{H.~L.}
  \bibnamefont{Swinney}}, \bibinfo{journal}{Phys. Rev. E}
  \textbf{\bibinfo{volume}{71}}, \bibinfo{pages}{030301(R)}
  (\bibinfo{year}{2005}).

\bibitem[{\citenamefont{Abate and Durian}(2006)}]{durian06}
\bibinfo{author}{\bibfnamefont{A.~R.} \bibnamefont{Abate}} \bibnamefont{and}
  \bibinfo{author}{\bibfnamefont{D.~J.} \bibnamefont{Durian}},
  \bibinfo{journal}{Phys. Rev. E} \textbf{\bibinfo{volume}{74}},
  \bibinfo{pages}{031308} (\bibinfo{year}{2006}).

\bibitem[{\citenamefont{van Noije et~al.}(1998)\citenamefont{van Noije, Ernst,
  and Brito}}]{noije1998}
\bibinfo{author}{\bibfnamefont{T.~P.~C.} \bibnamefont{van Noije}},
  \bibinfo{author}{\bibfnamefont{M.~H.} \bibnamefont{Ernst}}, \bibnamefont{and}
  \bibinfo{author}{\bibfnamefont{R.}~\bibnamefont{Brito}},
  \bibinfo{journal}{Phys. Rev. E.} \textbf{\bibinfo{volume}{57}},
  \bibinfo{pages}{R4891} (\bibinfo{year}{1998}).

\bibitem[{\citenamefont{van Noije et~al.}(1999)\citenamefont{van Noije, Ernst,
  Trizac, and Pagonabarraga}}]{noije1999}
\bibinfo{author}{\bibfnamefont{T.~P.~C.} \bibnamefont{van Noije}},
  \bibinfo{author}{\bibfnamefont{M.~H.} \bibnamefont{Ernst}},
  \bibinfo{author}{\bibfnamefont{E.}~\bibnamefont{Trizac}}, \bibnamefont{and}
  \bibinfo{author}{\bibfnamefont{I.}~\bibnamefont{Pagonabarraga}},
  \bibinfo{journal}{Phys. Rev. E} \textbf{\bibinfo{volume}{59}},
  \bibinfo{pages}{4326} (\bibinfo{year}{1999}).

\bibitem[{\citenamefont{Pagonabarraga et~al.}(2001)\citenamefont{Pagonabarraga,
  Trizac, van Noije, and Ernst}}]{pagonabarraga2001}
\bibinfo{author}{\bibfnamefont{I.}~\bibnamefont{Pagonabarraga}},
  \bibinfo{author}{\bibfnamefont{E.}~\bibnamefont{Trizac}},
  \bibinfo{author}{\bibfnamefont{T.~P.~C.} \bibnamefont{van Noije}},
  \bibnamefont{and} \bibinfo{author}{\bibfnamefont{M.~H.} \bibnamefont{Ernst}},
  \bibinfo{journal}{Phys. Rev. E.} \textbf{\bibinfo{volume}{65}},
  \bibinfo{pages}{011303} (\bibinfo{year}{2001}).

\bibitem[{\citenamefont{Fiege et~al.}(2009)\citenamefont{Fiege, Aspelmeier, and
  Zippelius}}]{fiege2009}
\bibinfo{author}{\bibfnamefont{A.}~\bibnamefont{Fiege}},
  \bibinfo{author}{\bibfnamefont{T.}~\bibnamefont{Aspelmeier}},
  \bibnamefont{and}
  \bibinfo{author}{\bibfnamefont{A.}~\bibnamefont{Zippelius}},
  \bibinfo{journal}{Phys. Rev. Lett.} \textbf{\bibinfo{volume}{102}},
  \bibinfo{pages}{098001} (\bibinfo{year}{2009}).

\bibitem[{\citenamefont{Kranz et~al.}(2010)\citenamefont{Kranz, Sperl, and
  Zippelius}}]{kranz2010}
\bibinfo{author}{\bibfnamefont{W.~T.} \bibnamefont{Kranz}},
  \bibinfo{author}{\bibfnamefont{M.}~\bibnamefont{Sperl}}, \bibnamefont{and}
  \bibinfo{author}{\bibfnamefont{A.}~\bibnamefont{Zippelius}},
  \bibinfo{journal}{Phys. Rev. Lett.} \textbf{\bibinfo{volume}{104}},
  \bibinfo{pages}{225701} (\bibinfo{year}{2010}).

\bibitem[{\citenamefont{Panaitescu and Kudrolli}(2010)}]{panaitescu2010}
\bibinfo{author}{\bibfnamefont{A.}~\bibnamefont{Panaitescu}} \bibnamefont{and}
  \bibinfo{author}{\bibfnamefont{A.}~\bibnamefont{Kudrolli}},
  \bibinfo{journal}{arXiv.org} \textbf{\bibinfo{volume}{arXiv:1001.0625v1
  [cond-mat.mtrl-sci]}} (\bibinfo{year}{2010}).

\bibitem[{\citenamefont{Williams and MacKintosh}(1996)}]{MacKintosh}
\bibinfo{author}{\bibfnamefont{D.~R.~M.} \bibnamefont{Williams}}
  \bibnamefont{and} \bibinfo{author}{\bibfnamefont{F.~C.}
  \bibnamefont{MacKintosh}}, \bibinfo{journal}{Phys. Rev. E}
  \textbf{\bibinfo{volume}{54}}, \bibinfo{pages}{R9} (\bibinfo{year}{1996}).

\bibitem[{\citenamefont{Espanol and Warren}(1995)}]{dissip_part}
\bibinfo{author}{\bibfnamefont{P.}~\bibnamefont{Espanol}} \bibnamefont{and}
  \bibinfo{author}{\bibfnamefont{P.}~\bibnamefont{Warren}},
  \bibinfo{journal}{Europhys. Lett.} \textbf{\bibinfo{volume}{30}},
  \bibinfo{pages}{191} (\bibinfo{year}{1995}).

\bibitem[{\citenamefont{Lubachevsky}(1991)}]{lubachevsky1991}
\bibinfo{author}{\bibfnamefont{B.~D.} \bibnamefont{Lubachevsky}},
  \bibinfo{journal}{J. Comput. Phys.} \textbf{\bibinfo{volume}{94}},
  \bibinfo{pages}{255} (\bibinfo{year}{1991}).

\bibitem[{\citenamefont{Reis et~al.}(2007)\citenamefont{Reis, Ingale, and
  Shattuck}}]{reis2007}
\bibinfo{author}{\bibfnamefont{P.~M.} \bibnamefont{Reis}},
  \bibinfo{author}{\bibfnamefont{R.~A.} \bibnamefont{Ingale}},
  \bibnamefont{and} \bibinfo{author}{\bibfnamefont{M.~D.}
  \bibnamefont{Shattuck}}, \bibinfo{journal}{Phys. Rev. Lett.}
  \textbf{\bibinfo{volume}{98}}, \bibinfo{pages}{188301}
  (\bibinfo{year}{2007}).

\bibitem[{\citenamefont{Garz\'{o} et~al.}(2007)\citenamefont{Garz\'{o}, Santos,
  and Montanero}}]{Garzo2007}
\bibinfo{author}{\bibfnamefont{V.}~\bibnamefont{Garz\'{o}}},
  \bibinfo{author}{\bibfnamefont{A.}~\bibnamefont{Santos}}, \bibnamefont{and}
  \bibinfo{author}{\bibfnamefont{J.~M.} \bibnamefont{Montanero}},
  \bibinfo{journal}{Physica A} \textbf{\bibinfo{volume}{376}},
  \bibinfo{pages}{94} (\bibinfo{year}{2007}).

\bibitem[{\citenamefont{Dufty et~al.}(1997)\citenamefont{Dufty, Brey, and
  Santos}}]{Dufty1997}
\bibinfo{author}{\bibfnamefont{J.~W.} \bibnamefont{Dufty}},
  \bibinfo{author}{\bibfnamefont{J.~J.} \bibnamefont{Brey}}, \bibnamefont{and}
  \bibinfo{author}{\bibfnamefont{A.}~\bibnamefont{Santos}},
  \bibinfo{journal}{Physica A} \textbf{\bibinfo{volume}{240}},
  \bibinfo{pages}{212} (\bibinfo{year}{1997}).

\bibitem[{\citenamefont{McNamara}(1993)}]{McNamara1993}
\bibinfo{author}{\bibfnamefont{S.}~\bibnamefont{McNamara}},
  \bibinfo{journal}{Phys. Fluids A} \textbf{\bibinfo{volume}{5}},
  \bibinfo{pages}{3056} (\bibinfo{year}{1993}).

\end{thebibliography}

\end{document}